\documentclass[12pt]{article}
\usepackage{amsmath,amssymb,bbm}
\usepackage{amsmath}
\begin{document}

\begin{center}

{\LARGE {Towards the core of the cosmological constant problem}}

\end{center}

\vskip 0.7cm
\begin{center}
{\large Eun Kyung Park\footnote{E-mail: ekpark1@dau.ac.kr}
and Pyung Seong Kwon\footnote{E-mail: bskwon@ks.ac.kr}\\
\vskip 0.2cm}
\end{center}

\vskip 0.05cm

\begin{center}
{\it $^1$Department of Materials Physics, Dong-A University,\\
Busan 604-714, Korea \\

\vskip 0.1cm

$^2$Department of Energy Science, Kyungsung University,\\
Busan 608-736, Korea}
\end{center}

\thispagestyle{empty}

\vskip 2.0cm
\begin{center}
{\bf Abstract}
\vskip 0.1cm
\end{center}

We apply a new self-tuning mechanism to the well-known Kachru-Kallosh-Linde-Trivedi (KKLT) model to address the cosmological constant problem. In this mechanism the cosmological constant $\lambda$ contains a supersymmetry breaking term ${\mathcal E}_{\rm SB}$ besides the usual scalar potential ${\mathcal V}_{\rm scalar}$ of the $N=1$ supergravity, which is distinguished from the usual theories where $\lambda$ is directly identified with ${\mathcal V}_{\rm scalar}$ alone. Also in this mechanism, whether $\lambda$ vanishes or not is basically determined by the tensor structure of the scalar potential density, not by the zero or nonzero values of the scalar potential itself. As a result of this application we find that the natural scenario for the vanishing $\lambda$ of the present universe is to take one of the AdS (rather than dS) vacua of KKLT as the background vacuum of our present universe. This AdS vacuum scenario has more nice properties as compared with dS vacua of the usual flux compctifications. The background vacuum is stable both classically and quantum mechanically (no tunneling instabilities), and the value $\lambda =0$ is also stable against quantum corrections because in this scenario the perturbative corrections of ${\mathcal V}_{\rm scalar}$ and quantum fluctuations $\delta_Q {\hat I}_{\rm brane}^{(NS)} + \delta_Q {\hat I}_{\rm brane}^{(R)}$ on the branes are all gauged away by an automatic cancelation between ${\mathcal V}_{\rm scalar} + \delta_Q {\hat I}_{\rm brane}^{(NS)} + \delta_Q {\hat I}_{\rm brane}^{(R)}$ and ${\mathcal E}_{\rm SB}$.

\vskip 3cm

\vskip 0.25cm
\begin{center}
{PACS number: 11.25.-w, 11.25.Uv}\\
\vskip 0.2cm
{\em Keywords}: cosmological constant problem, KKLT, supersymmetry breaking, self-tuning
\end{center}

\newpage
\setcounter{page}{1}
\setcounter{footnote}{0}

\baselineskip 6.0mm

\vskip 1cm
\hspace{-0.65cm}{\bf \Large I. Introduction}
\vskip 0.5cm
\setcounter{equation}{0}
\renewcommand{\theequation}{1.\arabic{equation}}

One of the most mysterious problems in the area of high energy physics including cosmology can be summarized as why the vacuum energy (or the cosmological constant) of our present universe is so small despite that the supersymmetry of our universe is considerably broken. Recently there has been proposed a new mechanism to address this cosmological constant problem in the framework of type IIB supergravity \cite{1}, where the four-dimensional cosmological constant $\lambda$ is forced to vanish by six-dimensional Einstein equation of the transverse sector, and therefore tunes itself to zero as a result. This mechanism is based on the viewpoint that our three-dimensional space is a stack of BPS (visible sector) $D3$-branes located at the conifold singularity of the Calabi-Yau threefold, and in this setup $\lambda$ generally appears as
\begin{equation}
\lambda = \frac{\kappa^2}{2} \Big( {\mathcal V}_{\rm scalar} + \delta_Q {\hat I}_{\rm brane}^{(NS)} + \delta_Q {\hat I}_{\rm brane}^{(R)} - {\mathcal E}_{\rm SB} \Big) \,\,.
\end{equation}
See Sec. 3.5 of this paper for the details.

In (1.1), ${\mathcal V}_{\rm scalar}$ is the usual scalar potential for the moduli of the $N=1$ supergravity, and $\delta_Q {\hat I}_{\rm brane}^{(NS)} + \delta_Q {\hat I}_{\rm brane}^{(R)}$ are NS-NS and R-R vacuum energies arising from quantum fluctuations (of the gravitational and standard model degrees of freedom with support) on the $D3$-branes. $\delta_Q {\hat I}_{\rm brane}^{(NS)}$ and $\delta_Q {\hat I}_{\rm brane}^{(R)}$ are expected to cancel out when supersymmetry of the brane region is unbroken. (The cancelation at one-loop order on the BPS $D3$-branes has been proven explicitly in Sec. VIIA of \cite{1} for the case where the three-form fluxes of type IIB theory are turned off.)  The last term ${\mathcal E}_{\rm SB}$ is a supersymmetry breaking term which originates from a gauge symmetry breaking of the R-R four-form $A_{(4)}$ arising at the quantum level in the brane region. So ${\mathcal E}_{\rm SB}$ is an energy scale of the supersymmetry breaking of the brane region, and at the same time it is also an energy scale of the gauge symmetry breaking (or an anomaly) generated by quantum fluctuations. Finally in (1.1), $\kappa^2 = 1/2 M_{pl}^2$ where $M_{pl}$ is the four-dimensional Planck scale. Eq. (1.1) is distinguished from the corresponding equation of the usual flux compactifications where $\lambda$ is simply given by ${\mathcal V}_{\rm scalar}$ alone. According to (1.1), $\lambda =0$ does not necessarily imply ${\mathcal V}_{\rm scalar} =0$ unlike in the usual $N=1$ supergravity, for instance, in \cite{1-1,2}.

${\mathcal E}_{\rm SB}$ is kind of an anomaly generated in the brane region, and for the $D3$-branes located at the conifold singularity of the Calabi-Yau threefold it takes (at one-loop order) the form
\begin{equation}
{\mathcal E}_{\rm SB}^{(1)} = -\delta_0 \int r^5 dr \epsilon_5 \rho_{\rm T}^{(1)} \,\,,~~~~~\big(\delta_0 = {\rm constant} \big) \,\,,
\end{equation}
(Compare (1.2) with (3.46), where $\rho_{\rm T}^{(1)}$ and $\delta \mu_{\rm T}^{m} (\phi)$ are given by (1.4) and (8.10), respectively) where $\epsilon_5 = \sqrt{det |{\hat h}_{mn}|} \, d\psi \wedge d\theta_1 \wedge d\phi_1 \wedge d\theta_2 \wedge d\phi_2$ is the volume-form of the base of the cone in the conifold metric $ds^2 = dr^2 + r^2 d \Sigma_{1,1}^2$ with
\begin{equation}
d \Sigma_{1,1}^2 = \frac{1}{9} \Big( d \psi + \sum_{i=1}^{2} \cos \theta_i d \phi_i \Big)^2 + \sum_{i=1}^{2} \frac{1}{6} \Big( d\theta_i^2 + \sin^2 \theta_i d\phi_i^2 \Big) \equiv {\hat h}_{mn} dy^m dy^n \,\,,
\end{equation}
so the volume of the base of the cone with unit radius is given by ${\rm Vol(B)}= \int \epsilon_5$. Also the integration $\int r^5 dr \epsilon_5$ in (1.2) is taken over the brane region, $0<r<r_B$, where $r_B$ being the thickness of the brane, and the constant $\delta_0$ is given by $\delta_0 = {6}/[{r_B^6 \,{\rm Vol(B)}}]$. (See Secs. VIB and VIIA of \cite{1} for the notations.) Finally $\rho_{\rm T}^{(1)}$ sources the supersymmetry breaking of the brane region (which is why ${\mathcal E}_{\rm SB}$ is called supersymmetry breaking term. See (8.11).), and it takes the form
\begin{equation}
\rho_{\rm T}^{(1)} (y)= \nu_{(1)}^m f_m(y) \,\,.
\end{equation}

In (1.4), $f_m (y)$'s are arbitrary gauge parameters and $\nu_{(1)}^m$'s represent (one-loop order) quantum excitations on the brane with components along the transverse directions of the $D3$-branes. Since $\rho_{\rm T}^{(1)}$ contains arbitrary gauge parameters, ${\mathcal E}_{\rm SB}$ in (1.2) also has gauge arbitrariness. In (1.1), ${\mathcal V}_{\rm scalar}$ takes nonzero values at the quantum level because it receives generically both perturbative and nonperturbative corrections. Also, $\delta_Q {\hat I}_{\rm brane}^{(NS)} + \delta_Q {\hat I}_{\rm brane}^{(R)}$ does not vanish if the brane supersummetry is broken. But $\lambda$ in (1.1) contains ${\mathcal E}_{\rm SB}$. So any nonzero ${\mathcal V}_{\rm scalar}$ and $\delta_Q {\hat I}_{\rm brane}^{(NS)} + \delta_Q {\hat I}_{\rm brane}^{(R)}$ can be gauged away by this ${\mathcal E}_{\rm SB}$ so that $\lambda$ vanishes as a result. Such a cancelation between ${\mathcal V}_{\rm scalar} + \delta_Q {\hat I}_{\rm brane}^{(NS)} + \delta_Q {\hat I}_{\rm brane}^{(R)}$ and ${\mathcal E}_{\rm SB}$ really occurs in (1.1), forced by a self-tuning equation (eq (3.41)) which imposes a constraint on $\lambda$. See Sec. 3.5 for this.

In the self-tuning mechanism of this paper, whether $\lambda$ vanishes or not is basically determined - in the six-dimensional internal space - by the tensor structure of the scalar potential density, not by the zero or nonzero values of the scalar potential ${\mathcal V}_{\rm scalar}$ itself. Thus in our self-tuning mechanism, whether ${\mathcal V}_{\rm scalar}$ vanishes or not is not important unlike in the usual theories where $\lambda=0$ is equivalent to ${\mathcal V}_{\rm scalar}=0$. In this paper we will apply this mechanism to the well-known scenario of KKLT \cite{2} to address the cosmological constant problem, especially aiming at explaining the (cause of the) vanishing cosmological constant of our present universe.

As a result of this application we find - basically in the framework of the type IIB $N=1$ supergravity - that the natural scenario for the vanishing $\lambda$ of our present universe is to take one of the AdS (rather than dS) vacua of KKLT as the background vacuum of the present universe. This AdS vacuum scenario has more nice properties as compared with dS vacua of the usual flux compactifications. The background vacuum is stable both classically and quantum mechanically (i.e., no tunneling instabilities), and the value $\lambda =0$ is perturbatively (radiatively) stable unlike in the usual theories because in our self-tuning mechanism of this paper the perturbative and nonperturbative corrections of ${\mathcal V}_{\rm scalar}$ are all gauged away by an automatic cancelation between ${\mathcal V}_{\rm scalar} + \delta_Q {\hat I}_{\rm brane}^{(NS)} + \delta_Q {\hat I}_{\rm brane}^{(R)}$ and ${\mathcal E}_{\rm SB}$.

\vskip 1cm
\hspace{-0.65cm}{\bf \Large II. Scalar potential of KKLT}
\vskip 0.5cm
\setcounter{equation}{0}
\renewcommand{\theequation}{2.\arabic{equation}}

Kachru $et~al.$ have shown in the framework of the Klebanov-Strassler (KS) compactifications \cite{3} that one can construct a de Sitter (dS) vacuum (of type IIB theory) with broken supersymmetry if we allow for nonperturbative corrections and anti-$D3$-branes. They first obtained a supersymmetric anti-de Sitter (AdS) vacuum from the superpotential of the form
\begin{equation}
W=W_0 + A e^{i a \rho} \,\,,
\end{equation}
where $W_0$ is a tree level contribution arising from the fluxes and does not contain the K$\ddot{\rm a}$hler modulus $\rho$. The second term is a nonperturbative correction coming from Euclidean $D3$-branes (instantons) \cite{4}, or the gaugino condensation in the $N=1$ supersymmetric $SU(N_c)$ gauge theory generated by the stack of $N_c$ coincident $D7$-branes wrapping four-cycles in the Calabi-Yau threefold \cite{5}. Since $W$ contains K$\ddot{\rm a}$hler modulus the no-scale structure of the Lagrangian has been broken and the supersymmetric vacuum is now described by
\begin{equation}
D W=0 \,\,,
\end{equation}
but not necessarily $W=0$, where the covariant derivative $D_a W$ is defined by $D_a W= \partial_a W + (\partial_a {\mathcal K}) W$, and where the K${\ddot{\rm a}}$hler potential ${\mathcal K}$ is given at the tree level of type IIB by (see \cite{6})
\begin{equation}
{\mathcal K}=-3 \ln \big[-i (\rho - {\bar \rho}) \big] - \ln \big[-i (\tau - {\bar \tau}) \big] - \ln \big[-i \int_{\mathcal M_6} \Omega \wedge {\bar \Omega} \,\big]\,\,.
\end{equation}
In (2.3), $\tau$ is type IIB axion/dilaton and $\Omega$ is holomorphic three-form of the Calabi-Yau threefold ${\mathcal M}_6$.

From the superpotential $W$ and the K${\ddot{\rm a}}$hler potential $\mathcal K$ one can construct the scalar potential of the ${\mathcal N}=1$ supergravity \cite{1-1,6}:
\begin{equation}
{\mathcal V}_{\rm scalar} = \frac{1}{2 \kappa^2_{10}} e^{\mathcal K} \Big(\, {\mathcal G}^{a {\bar b}} D_a W \,\overline{D_b W} -3 |W|^2 \,\Big) \,\,,
\end{equation}
where ${\mathcal G}_{a {\bar b}}=\partial_a \partial_{\bar b} {\mathcal K}$, and $a$, $b$ are summed all over the complex structure moduli $\tau^I$, the axion/dilaton $\tau$ and the K${\ddot{\rm a}}$hler modulus $\rho$. For the no-scale structure \cite{6,7} in which $W=W_0$, (2.4) reduces to
\begin{equation}
{\mathcal V}_{\rm scalar} = \frac{1}{2 \kappa^2_{10}} e^{\mathcal K} \Big(\, {\mathcal G}^{i {\bar j}} D_i W_0 \,\overline{D_j W_0} \Big) \,\,,
\end{equation}
where $i$, $j$ are now summed over $\tau^I$ and $\tau$, and the superpotential $W_0$ is given by
\begin{equation}
W_0 = \int_{\mathcal M_6} G_{(3)} \wedge \Omega \,\,, ~~~~~~~~~~ \big( G_{(3)} = F_{(3)} - \tau H_{(3)} \big)\,\,,
\end{equation}
where $F_{(3)}$ and $H_{(3)}$ are R-R resp. NS-NS three-form field strengths. If we take $F_{(3)}$ and $H_{(3)}$ as $F_{(3)}$, $H_{(3)}$ $\in H^3 ({\mathcal M}_6 , {\mathcal Z})$, then the potential (2.5) fixes the moduli at values for which $G_{(3)}$ is imaginary self-dual (ISD) at the tree level \cite{2}. But once the nonperturbative term comes in, $G_{(3)}$ will not be ISD anymore. Concerning this point a little more explanation may be necessary as follows.

In the original KKLT the authors used a two step procedure in which they first fix the complex structure moduli (and also the dilaton moduli as well) at values where $G_{(3)}$ becomes ISD, and then fix the K$\ddot{\rm a}$hler modulus by introducing nonperturbative correction to the superpotential (see (2.1)) in such a way that $G_{(3)}$ still remains ISD. This is possible if the masses of the complex structure moduli and the dilaton moduli are much larger than the mass of the K$\ddot{\rm a}$hler modulus. Indeed in KKLT the complex structure moduli are fixed at string scale and they are integrated out. Hence in KKLT the instanton determinant $A$ in (2.1) is effectively a constant and the shifts of the complex structure moduli from their classical ISD positions are consequently negligible.

But after this original KKLT, there also came out some other articles in which the KKLT mechanism of moduli stabilization is extended to more general cases where the complex structure (and the dilaton) moduli are not integrated out anymore and hence they appear explicitly in the effective theory \cite{8,9}. In these theories the instanton determinant depends on the complex structure moduli and $G_{(3)}$ now acquires imaginary anti self-dual (IASD) components by the nonperturbative corrections. In our present paper we want to extend our discussions as much as possible so that the self-tuning mechanism of this paper can be applied even to these generalized theories as well. Hence in our paper we will include these IASD components when we investigate the whole possible contributions to the scalar potentials of the background vacua. We will be back to this point later. (See for instance the paragraph below eq. (4.3).)

Turning back to the superpotential (2.1), one can concentrate only on the K${\ddot{\rm a}}$hler modulus $\rho$ if we neglect the no-scale part (2.5) (for a moment) as in the original KKLT. (But see the first paragraph of Sec. 4.1.) The scalar potential is therefore given by
\begin{equation}
{\mathcal V}_{\rm scalar} = \frac{1}{2 \kappa^2_{10}} e^{\mathcal K} \Big(\, {\mathcal G}^{\rho {\bar \rho}} D_{\rho} W \,\overline{D_{\rho} W} -3 |W|^2 \,\Big) \,\,,
\end{equation}
and using (2.1) one obtains \cite{2}
\begin{equation}
{\mathcal V}_{\rm scalar} = \frac{1}{2 \kappa^2_{10}}\, e^{{\mathcal K}_{\tau}+{\mathcal K}_{\rm cs}}\Bigg[ \frac{a A e^{-a \sigma}}{2 \sigma^2} \Big(\, \frac{1}{3} a A \sigma e^{-a \sigma} + W_0 +  A e^{-a \sigma} \Big) \Bigg] \,\,,
\end{equation}
where the axion in $\rho$ has been set to zero and $\sigma$ is defined by $\rho = i\sigma$. From (2.7) one finds that the minimum of ${\mathcal V}_{\rm scalar}$ takes negative values for the superpotential $D_a W =0$, and therefore it describes supersymmetric AdS vacua because ${\mathcal V}_{\rm scalar}$ is identified with the four-dimensional cosmological constant $\lambda$ in the usual flux compactifications including KKLT.

At the final step of KKLT the AdS minimum is uplifted to a dS minimum by the anti-${D3}$-branes introduced at the tip of the KS throat where the introduction of anti-${D3}$-branes does not violate the tadpole condition. By this process ${\mathcal V}_{\rm scalar}$ acquires an additional term\footnote{Our convention is $\rho = \frac{b}{\sqrt 2} + i e^{4u}$ and the prefactor $\frac{1}{2 \kappa_{10}^2} e^{{\mathcal K}_{\tau}+{\mathcal K}_{\rm cs}}$ of (2.8) has been omitted in (2.9).}
\begin{equation}
\delta {\mathcal V}_{\rm scalar} = \frac{D}{\sigma^2} \,\,,
\end{equation}
where $D$ is a positive constant proportional to the number of anti-${D3}$-branes. So in \cite{2} Kachru $et~al.$ obtain dS vacua by fine-tuning the constant $D$ so that the minimum of the resulting ${\mathcal V}_{\rm scalar}$ becomes very close to zero.

\vskip 1cm
\hspace{-0.65cm}{\bf \Large III. A self-tuning mechanism for $\bold \lambda$}
\vskip 0.5cm
\setcounter{equation}{0}
\renewcommand{\theequation}{3.\arabic{equation}}

In general the dS vacua uplifted by anti-$D3$-branes can have two different kinds of tunneling instabilities (see Sec. 5.2), one of which is related to the fact that $\lambda$ is directly given by ${\mathcal V}_{\rm scalar}$ alone in these theories. The scalar potential ${\mathcal V}_{\rm dS}$ of the dS minimum at $\sigma =\sigma_m$ takes positive (though it is very small) values, while ${\mathcal V}_{\rm scalar}$ asymptotically vanishes, ${\mathcal V}_{\rm scalar}|_{\sigma \rightarrow \infty}=0$. So these dS vacua are only local minima of the potential which eventually decay into the run away vacuum at $\sigma =\infty$ which corresponds to a Minkowski space with a large (or a decompactified) internal Calabi-Yau volume.

In the original KKLT, however, it was shown that the lifetime of the dS vacua is larger than the cosmological time scale of $10^{10}$ years in certain approximations. So KKLT does not suffer from this tunneling instability problem. Besides this, the elegance of KKLT is that all stringy corrections are very small. Both $g_s$- and $\alpha^{\prime}$-corrections are small in the part of moduli space their vacuum lives and hence the quantum corrections are only subleading.

But still, though the KKLT is an attractive scenario for the late-time cosmology with a small positive cosmological constant, it has some difficulties as for being a realistic model of our universe, especially when looking at from a standpoint of the cosmological constant problem. In this paper we propose a new self-tuning mechanism in which the fine-tuning $\lambda=0$ is automatically achieved by a certain constraint (or a self-tuning) equation. As shown in (1.1) $\lambda$ contains an extraordinary term ${\mathcal E}_{\rm SB}$ which possesses gauge arbitrariness, and the whole quantum fluctuations $\delta_Q {\hat I}_{\rm brane}^{(NS)} + \delta_Q {\hat I}_{\rm brane}^{(R)}$ on the branes and nonzero contributions to ${\mathcal V}_{\rm scalar}$ coming from perturbative and nonperturbative corrections are all gauged away by ${\mathcal E}_{\rm SB}$ (and by a self-tuning equation as mentioned above) and as a result the fine-tuning $\lambda =0$ is always preserved.

Beside this, the instabilities of the background vacua are innately absent in our case. As described above $\lambda$ contains an additional term ${\mathcal E}_{\rm SB}$, and hence in (1.1) we are allowed to take negative values for ${\mathcal V}_{\rm scalar}$ at the minimum $\sigma =\sigma_m$ because in our case ${\mathcal V}_{\rm scalar} <0$ does not directly imply a negative $\lambda$ due to the presence of this ${\mathcal E}_{\rm SB}$. Indeed in our self-tuning mechanism the background state of our present universe will be identified with one of the AdS (rather than dS) vacua of KKLT. (See the AdS vacuum scenario proposed in Sec. 5.2.) So the instabilities of dS vacua described above are essentially irrelevant to our case. In this section we will discuss about the basic principle of our self-tuning mechanism described above, together with brief reviews of some formulas and ideas presented in \cite{1} if necessary for reader's convenience.

\vskip 0.3cm
\hspace{-0.6cm}{\bf \large 3.1 Six-dimensional Einstein equation}
\vskip 0.15cm

In the string frame the type IIB action is given by
\begin{equation}
I_{\rm IIB} = \frac{1}{2 \kappa_{10}^2} \int d^{10} x \sqrt{-G} \Big\{ e^{-2 \phi} \big[ {\mathcal R}_{10} + 4 (\nabla \phi)^2 \big] - \frac{1}{2} F_{(1)}^2 - \frac{1}{2 \cdot 3!} G_{(3)} \cdot {\bar G}_{(3)} - \frac{1}{4 \cdot 5!} {\tilde F}_{(5)}^2 \Big\} \nonumber
\end{equation}
\begin{equation}
+ \frac{1}{8 i \kappa_{10}^2} \int e^{\phi} A_{(4)} \wedge G_{(3)} \wedge {\bar G}_{(3)} \,\,,
\end{equation}
where $\phi$ is the dilaton with $e^{\phi} = g_s e^{\hat{\phi}}$, and $F_{(1)} =dA_{(0)}$, ${\tilde F}_{(5)} =F_{(5)} -\frac{1}{2} A_{(2)} \wedge H_{(3)} + \frac{1}{2} B_{(2)} \wedge F_{(3)}$ with $F_{(n+1)} =dA_{(n)}$. Among these field strengths, ${\tilde F}_{(5)}$ is self-dual and the ansatz is given by
\begin{equation}
{\tilde F}_{(5)} = (1+ \ast_{10}) d \xi (y) \wedge \sqrt{-g_4}\,\, dx^0 \wedge dx^1 \wedge dx^2 \wedge dx^3 \,\,.
\end{equation}
In addition to this we have the local terms
\begin{equation}
I_{\rm brane} = - \int d^4 x \sqrt{-det \,(G_{\mu\nu})}\,\, T(\phi) + \mu (\phi) \int A_{(4)} \,\,,
\end{equation}
where $G_{\mu\nu}$ is a pullback of the target space metric $G_{MN}$ to the four-dimensional brane world. Also $T(\phi)$ represents the tension of the $D3$-brane and at the tree level it is given by $T(\phi) = T_3 e^{-\phi}$. But at the quantum level it becomes $T(\phi) = T_3 e^{-\phi} + \rho_{\rm vac} (\phi)$, where $\rho_{\rm vac} (\phi)$ represents quantum correction terms (see for instance ref. \cite{10}) and it is identified with NS-NS sector vacuum energy density of the three-dimensional space. Similarly, $\mu (\phi)$ is simply $\mu (\phi) = \mu_3$ at the tree level. But it turns into $\mu (\phi)=\mu_3 + \delta \mu (\phi)$ at the quantum level, where $\delta \mu (\phi)$ is an R-R counterpart of $\rho_{\rm vac} (\phi)$ representing R-R sector vacuum energy density of the three-dimensional space.

Upon reduction (see Sec. IV of \cite{1})
\begin{equation}
ds_{10}^2 = e^{B(y)} g_{\mu\nu}(x) dx^{\mu} dx^{\nu} + e^{{\hat \phi}(y) - B(y)} h_{mn}(y) dy^m dy^n \,\,,
\end{equation}
where $\mu,\nu =(0,1,2,3)$, $m,n =(5, \cdots , 10)$, the type IIB action (3.1) reduces into
\begin{equation}
I_{\rm IIB} = \frac{1}{2 \kappa_{10}^2 g_s^2} \Big( \int d^4 x \sqrt{-g_4} {\mathcal R}_4 (g_{\mu\nu}) \, \Big)\Big(\int d^6 y \,\sqrt{h_6} e^{{\hat \phi} - 2B} \,\Big) + \frac{1}{2 \kappa_{10}^2 g_s^2} \Big( \int d^4 x \sqrt{-g_4} \,\Big) \nonumber
\end{equation}
\begin{equation}
\times \Big( \int d^6 y \,\sqrt{h_6} \, \big( {\mathcal R}_6 (h_{mn}) - {\mathcal L}_F \big) \Big) + {\rm topological~\,\, term} \,\,,~~~~~~~~~
\end{equation}
where ${\mathcal L}_F$ is given by
\begin{equation}
{\mathcal L}_F = (\partial \hat{\phi})^2 - 2(\partial \hat{\phi})(\partial B) + 2 (\partial B)^2 + \frac{g_s^2}{2} e^{2 \hat{\phi}} (\partial A_{(0)} )^2 - \frac{g_s^2}{2} e^{2 \hat{\phi} -4B} (\partial {\xi}\,)^2 \nonumber
\end{equation}
\begin{equation}
+ \frac{g_s^2}{2 \cdot 3!} e^{2B} G_{mnp} \bar{G}^{mnp} \,\,,~~~~~~~~~~~~~~~~~~~~
\end{equation}
(But see also the sentences below eq. (3.17). At the quantum level the Lagrangian ${\mathcal L}_F$ can also include off-shell contributions coming from perturbative and nonperturbative corrections. See Sec. VII as an example.) and the topological term comes from the Chern-Simons term $\int e^{\phi} A_{(4)} \wedge G_{(3)} \wedge \bar{G}_{(3)}$, which does not involve any moduli (except the dilaton $\tau$) or the metric. From (3.5) the six-dimensional action defined on the internal space can be written as
\begin{equation}
I_{\rm IIB}/\Big(\int d^4 x \sqrt{-g_4}\Big)  = \frac{1}{2 \kappa_{10}^2 g_s^2}  \int d^6 y \,\sqrt{h_6} \, \Big( {\mathcal R}_6 (h_{mn}) - {\mathcal L}_F + \beta e^{{\hat \phi} - 2B} \,\Big) + {\rm topological~term}\,\,,
\end{equation}
where $\beta$ is defined by
\begin{equation}
\beta = \frac{\int d^4 x \sqrt{- g_4} {\mathcal R}_4 (g_{\mu\nu})}{\int d^4 x \sqrt{- g_4} } \,\,,
\end{equation}
and hence on the brane, $\beta= 4 \lambda$ for the maximally symmetric spacetime. Varying (3.7) with respect to $\delta h^{mn}$ one obtains
\begin{equation}
\mathcal R_{mn} - \frac{1}{2} h_{mn} \mathcal R_6 - \frac{1}{2} T_{mn} - \frac{\beta}{2}  e^{{\hat\phi}-2B} h_{mn}=0 \,\,,
\end{equation}
where the energy momentum tensor $T_{mn}$ is defined by
\begin{equation}
T_{mn} = \frac{2}{\sqrt{h_6}} \frac{\delta I_F}{\delta h^{mn}}\,\,, ~~~~~~\big( I_F \equiv \int d^6 y \sqrt{h_6} {\mathcal L}_F \big) \,\,.
\end{equation}
(3.9) does not involve local terms arising from (3.3) because $D3$-branes do not couple to the unwarped metric $h_{mn}$ in the action (3.3).

In (3.9), $\mathcal R_{mn}$ and $\mathcal R_6$ vanish at the classical level because the internal Calabi-Yau is Ricci-flat. But at the quantum level, $h_{mn}$ acquires correction terms in the equations of motion,
\begin{equation}
h_{mn} = h_{mn}^{(0)} +  h_{mn}^{(1)} + h_{mn}^{(2)} + \cdots \,= h_{mn}^{(0)} + \delta_Q h_{mn} \,\,,
\end{equation}
in our perturbation scheme (see (7.13)). Hence in (3.9) (and also in what follows) we can not take $\mathcal R_{mn} (h_{mn}) = \mathcal R_6  (h_{mn})=0$ at the quantum level though we have $\mathcal R_{mn} (h_{mn}^{(0)}) = \mathcal R_6 (h_{mn}^{(0)}) = 0$, because they do not vanish at off-shell. Besides the perturbations, there are also backreactions of the fluxes and local sources like $D3$-branes which carry standard model fields etc. These backreactions on the internal geometry also could yield $R_{mn} \neq 0$ and $\mathcal R_6 \neq 0$. In this paper such  deformations of internal geometry caused by perturbations and backreactions are all under consideration because we never set $\mathcal R_{mn} = \mathcal R_6 =0$ in the whole procedure of our discussions as mentioned above.\footnote{There is another viewpoint on this backreaction problem. For instance in Sec. III of ref. [1] (also see ref. [6] therein) it was argued that the Calabi-Yau three-folds may be thought of as NS-NS solitons whose ADM masses are proportional to $1/ g_s^2$. Hence in the limit $g_s \rightarrow 0$ these Calabi-Yau three-folds are very heavy and rigid, and consequently deformations of internal geometry due to backreactions are highly suppressed.} The nonzero $\mathcal R_{mn}$ and $\mathcal R_6$ cancel out themselves during the process of obtaining the self-tuning equation (3.30).

\vskip 0.3cm
\hspace{-0.6cm}{\bf \large 3.2 Four-dimensional cosmological constant $\lambda$}
\vskip 0.15cm

The four-dimensional action defined on the external space can be obtained by rewriting (3.5) as
\begin{equation}
I_{\rm IIB} = \frac{1}{2\kappa^2} \int d^4 x \sqrt{-g_4} {\mathcal R}_4 (g_{\mu\nu}) + \int d^4 x \sqrt{-g_4} {\hat I}_{\rm bulk} + {\rm topological~term}\,\,,
\end{equation}
where ${2\kappa^2} \equiv 2 \kappa_{10}^2 g_s^2 / \big( \int d^6 y \,\sqrt{h_6} e^{{\hat \phi} - 2B}\big)$ and ${\hat I}_{\rm bulk}$ is defined by
\begin{equation}
 {\hat I}_{\rm bulk} = \frac{1}{2 \kappa_{10}^2 g_s^2}  \int d^6 y \,\sqrt{h_6} \big( {\mathcal R}_6 (h_{mn}) - {\mathcal L}_F \big) \,\,.
\end{equation}
Adding (3.3) to (3.12) one can show that the total action $I_{\rm IIB} + I_{\rm brane}$ can be written in the form
\begin{equation}
I_{\rm total} = \frac{1}{2\kappa^2} \int d^4 x \sqrt{-g_4} \big( {\mathcal R}_4 (g_{\mu\nu}) - 2 \lambda \big) + {\rm topological~term}\,\,,
\end{equation}
where the cosmological constant $\lambda$ is defined by
\begin{equation}
\lambda = - \kappa^2 [ {\hat I}_{\rm bulk} + {\hat I}_{\rm brane}] \,\,,
\end{equation}
where ${\hat I}_{\rm brane} \equiv I_{\rm brane}/ \Big(\int d^4 x \sqrt{-g_4}\Big)$.

Turning back to (3.6) we see that the Lagrangian ${\mathcal L}_F$ can be written as
\begin{equation}
{\mathcal L}_F = K - V\,\,, ~~~~~~~~~~(\,K=h^{mn}K_{mn}\,) \,\,,
\end{equation}
where $K_{mn}$ and $V$, the kinetic and potential parts of the Lagrangian, take respectively the forms $K_{mn} = \sum_{A,B} F_{AB} (\phi_C ) \partial_{m} \phi_A \partial_{n} \phi_B$ and $V=V(\phi_A , h^{mn})$, where $\phi_A$'s represent the six-dimensional scalar fields such as $\hat{\phi}$, $B$, $A_{(0)}$ and $\xi$ etc. Namely in (3.16), while $V$ involves $h^{mn}$, $K_{mn}$ does not. Also in (3.16), $V$ is related to ${\mathcal V}_{\rm scalar}$ by the equation
\begin{equation}
{\mathcal V}_{\rm scalar} = \frac{1}{2 \kappa_{10}^2 g_s^2} \int d^6 y \sqrt{h_6} V \,\,,
\end{equation}
and thus for the no-scale structure the potential density $V$ arising from the fluxes is identified with $-\frac{g_s^2}{3!} G_{(3)}^{\rm IASD} \cdot {\bar G}_{(3)}^{\rm IASD}$ (see (4.3)) in the case of type IIB action. But in general, $V$ also includes the off-shell contributions coming from perturbative and nonperturbative corrections including those, for instance the $D3$-brane potential induced by IASD fluxes in Sec VII etc.

Now we substitute (3.10) - with ${\mathcal L}_F$ given by (3.16) - into (3.9) and contract the indices $m$ and $n$. Then we obtain
\begin{equation}
{\mathcal R}_6 - {\mathcal L}_F - \frac{1}{2} ({\mathcal N}-1) V +\frac{3}{2} \beta e^{{\hat \phi} - 2B} =0 \,\,,
\end{equation}
where ${\mathcal N}$ is defined by ${\mathcal N}\equiv h^{mn} \frac{\partial}{\partial h^{mn}}$. Again, we do not take ${\mathcal R}_6 =0$ in (3.18) because $h_{mn}$ in ${\mathcal R}_6$ (and other fields in (3.18) as well) contains correction terms coming from perturbatons. But integrating (3.18) and using (3.13) one finds that
\begin{equation}
{\hat I}_{\rm bulk}= - \frac{3 \beta}{4 \kappa^2} + \frac{1}{4\kappa_{10}^2 g_s^2} \int d^6 y \,\sqrt{h_6} \big( {\mathcal N}-1 \big) V \,\,,
\end{equation}
and substituting (3.19) into (3.15) (and using $\beta = 4 \lambda$) one finally obtains
\begin{equation}
\lambda = \frac{\kappa^2}{8\kappa_{10}^2 g_s^2} \int d^6 y \,\sqrt{h_6} \big( {\mathcal N}-1 \big) V  + \frac{\kappa^2}{2} {\hat I}_{\rm brane} \,\,,
\end{equation}
which is now independent of ${\mathcal R}_6 (h_{mn})$.

\vskip 0.3cm
\hspace{-0.6cm}{\bf \large 3.3 Self-tuning equation for $\lambda$}
\vskip 0.15cm

Now we proceed to obtain a self-tuning equation for $\lambda$, which is one of the main points of this paper. First, we substitute ${\mathcal L}_F$ in (3.18) into (3.10) to get
\begin{equation}
T_{mn} = 2 ({\mathcal R}_{mn} - \frac{1}{2} h_{mn} {\mathcal R}_{6}) + \frac{1}{2} h_{mn} ({\mathcal N} - 1) V - \frac{\partial}{\partial h^{mn}}({\mathcal N} - 1) V  - \frac{3}{2} \beta e^{{\hat \phi}-2B} h_{mn} \,\,.
\end{equation}
Next, substitute (3.21) into (3.9) and contract $m$ and $n$. Then we obtain
\begin{equation}
\beta = - \frac{1}{3} \chi^{1/2} ({\mathcal N} - 1)({\mathcal N} -3) V \,\,,~~~~~~\big( \chi^{1/2} \equiv e^{2B -{\hat \phi}} \big)\,\,.
\end{equation}
Let us repeat the same procedure again. Substitute (3.22) into (3.18) to obtain
\begin{equation}
{\mathcal L}_F = {\mathcal R}_6 - \frac{1}{2} ({\mathcal N} - 1 )({\mathcal N} -2) V \,\,.
\end{equation}
Next, substitute (3.23) into (3.10) to obtain
\begin{equation}
T_{mn} = 2 \big( \mathcal R_{mn} - \frac{1}{2} h_{mn} \mathcal R_6 \big) + \frac{1}{2} h_{mn} ({\mathcal N} - 1)({\mathcal N}- 2)V
- \frac{\partial}{\partial h^{mn}} ({\mathcal N} - 1)({\mathcal N} - 2) V \,\,.
\end{equation}
Finally, substitute (3.24) back into (3.9) and contract $m$ and $n$. We obtain
\begin{equation}
\beta = \frac{1}{6} \, \chi^{1/2} ({\mathcal N} - 1)({\mathcal N} -2)({\mathcal N} -3) V \,\,.
\end{equation}

(3.22) and (3.25) suggest that $\beta$ always contains the operators $({\mathcal N} -1)$ and $({\mathcal N} -3)$ in common. We prove this as follows. First, we observe that $\beta$'s in (3.22) and (3.25) both take the form
\begin{equation}
\beta = b_0 \chi^{1/2} ({\mathcal N} -1) \Pi ({\mathcal N}) V \,\,,
\end{equation}
where $b_0$ is a constant and $\Pi ({\mathcal N})$ is an operator of the form
\begin{equation}
\Pi ({\mathcal N}) = \sum_{k} c_k ({\mathcal N} -n_1 ) \cdots ({\mathcal N} -n_k ) \,\,,
\end{equation}
where $n_i$ are integers. So we start by assuming that $\beta$ always appears in the form (3.26). Now we substitute (3.26) into (3.18) to obtain
\begin{equation}
{\mathcal L}_F = {\mathcal R}_6 - \frac{1}{2} ({\mathcal N} - 1) (1-3 b_0 \Pi ({\mathcal N})) V \,\,.
\end{equation}
Next, substitute (3.28) into (3.10) to obtain
\begin{equation}
T_{mn} = 2 \big({\mathcal R}_{mn} -\frac{1}{2} h_{mn} {\mathcal R}_6 \big) + \frac{1}{2} h_{mn}  ({\mathcal N} - 1) (1-3 b_0 \Pi ({\mathcal N})) V - \frac{\partial}{\partial h^{mn}} ({\mathcal N} - 1) (1-3 b_0 \Pi ({\mathcal N})) V \,\,.
\end{equation}
Finally, substitute (3.29) back into (3.9) and contract $m$ and $n$. Then we obtain
\begin{equation}
\beta = \frac{1}{6} \chi^{1/2} ({\mathcal N} -1) ({\mathcal N} -3) (1-3 b_0 \Pi ({\mathcal N})) V \,\,.
\end{equation}
(3.30) takes the form (3.26) again, which ensures that the prerequisite assumption (3.26) on $\beta$ is valid. Also (3.30) shows that $\beta$ always contains $({\mathcal N} -1)$ and $({\mathcal N} -3)$ acting on $V$, which proves the proposition.

We obtained (3.30) starting from the Einstein equation (3.9). But (3.30) does not contain ${\mathcal R}_6$ of the perturbed $h_{mn}$ because it has canceled out during the process of obtaining (3.30). This suggests that $\lambda$ (Recall that $\beta = 4\lambda$) is not affected by the deformations of internal geometry caused by quantum fluctuations (and backreactions which also could yield ${\mathcal R}_{mn} \neq 0$ and ${\mathcal R}_6 \neq 0$) at least in the supergravity framework. Besides this, (3.30) suggests a very important fact. According to (3.30), whether $\lambda$ vanishes or not is entirely determined by the tensor structure of $V$, not by any other factors like zero or nonzero values of the scalar potential ${\mathcal V}_{\rm scalar}$ etc. We will be back to this point in Sec. 3.5.

\vskip 0.3cm
\hspace{-0.6cm}{\bf \large 3.4 Gauge symmetry breaking}
\vskip 0.15cm

\vskip 0.3cm
\hspace{-0.6cm}{(1) $I_{\rm brane}^{(R)}$ at the quantum level}
\vskip 0.15cm

(3.3) shows that $I_{\rm brane}$ consists of two (NS-NS and R-R) parts. Among these two, the second part represents an electric coupling of $D3$-branes to the R-R four-form $A_{(4)}$ and it can be rewritten as
\begin{equation}
I_{\rm brane}^{(R)} = \frac{1}{4!} \int d^4 x A_{\mu_0 \mu_1 \mu_2 \mu_3} J^{\mu_0 \mu_1 \mu_2 \mu_3} \,\,,
\end{equation}
where $J^{\mu_0 \mu_1 \mu_2 \mu_3}$ is the world volume current density of the $D3$-brane,
\begin{equation}
J^{\mu_0 \mu_1 \mu_2 \mu_3} = \mu_3 \epsilon^{\alpha_0 \alpha_1 \alpha_2 \alpha_3} \big(\frac{\partial X^{\mu_0}}{\partial x^{\alpha_0}} \big) \cdots \big(\frac{\partial X^{\mu_3}}{\partial x^{\alpha_3}} \big) \,\,.
\end{equation}
At the classical level $J^{\mu_0 \mu_1 \mu_2 \mu_3}$ is just a solitonic current density, $J^{\mu_0 \mu_1 \mu_2 \mu_3}_{\rm sol}$, representing classical world volume dynamics of the $D3$-brane. In that case $X^{\mu} (x)$'s in (3.32) stand for the classical fields, $X^{\mu}_{\rm cl} (x)$, defined on the world volume of the $D3$-brane, and for the embedding $X^{\mu}_{\rm cl} (x) = x^{\mu}$, $J^{0123}_{\rm sol}$ is simply $\mu_3$. At the quantum level, however, $X^{\mu}(x)$'s include fluctuations $X^{\mu^{\prime}}$, $X^{\mu} = X^{\mu}_{\rm cl} + X^{\mu^{\prime}}$.

Since $X^{\mu^{\prime}}$'s are fluctuations of the open string degrees of freedom, they correspond to the fluctuations of the standard model fields with support on the $D3$-brane. Due to these fluctuations $J^{\mu_0 \mu_1 \mu_2 \mu_3}$ acquires an additional term, $J^{\mu_0 \mu_1 \mu_2 \mu_3} = J^{\mu_0 \mu_1 \mu_2 \mu_3}_{\rm sol} + <\chi^{\mu_0 \mu_1 \mu_2 \mu_3}_{\rm vac}>$, where $<\chi^{\mu_0 \mu_1 \mu_2 \mu_3}_{\rm vac}>$ represents quantum corrections corresponding to the fluctuations (of the standard model degrees of freedom with support) on the $D3$-brane. Denoting $J^{0 1 2 3}_{\rm sol}$ and $<\chi^{0 1 2 3}_{\rm vac}>$, respectively, by $\mu_3$ and $\delta \mu (\phi)$, one can rewrite (3.31) as
\begin{equation}
I_{\rm brane}^{(R)} = \big[ \int d^4 x \sqrt{-g_4} \,\big] \, \int d^6 y \sqrt{h_6} \,\, \mu (\phi) \xi(y) \delta^6 (y) \,\,,
\end{equation}
where we have used
\begin{equation}
A_{(4)} = \xi(y) \sqrt{-g_4} \,\, dx^0 \wedge dx^1 \wedge dx^2 \wedge dx^3 \,\,,
\end{equation}
and the normalization convention $\int d^6 y \sqrt{h_6} \,\, \delta^6 (y) =1$ of the six-dimensional delta function.
(3.33) coincides with the second term of (3.3), and where $\mu(\phi) = \mu_3 + \delta \mu (\phi)$ as before.

\vskip 0.3cm
\hspace{-0.6cm}{(2) Gauge symmetry breaking}
\vskip 0.15cm

Going back to the classical level, the second term of (3.3) is invariant under the gauge transformation $A_{(4)} \rightarrow A_{(4)} + \delta A_{(4)}$ with $\delta A_{(4)} = d \Lambda_{(3)}$, where $\Lambda_{(3)}$ is an arbitrary three-form. Indeed $\delta_G I_{\rm brane}^{(R)}$ vanishes for $\delta A_{(4)} = d \Lambda_{(3)}$ : $\delta_G I_{\rm brane}^{(R)} = \mu_3 \int_{\partial \Sigma} \Lambda_{(3)} =0$ because $\Lambda_{(3)}$ is assumed to vanish at the boundary $\partial \Sigma$ of the four-dimensional spacetime. But once we go up to quantum level, $I_{\rm brane}^{(R)}$ is not gauge invariant anymore. The reason is because while $J^{\mu_0 \mu_1 \mu_2 \mu_3}_{\rm sol}$ satisfies $\partial_{\mu_0} J^{\mu_0 \mu_1 \mu_2 \mu_3}_{\rm sol}=0$, the off-shell quantity $<\chi^{\mu_0 \mu_1 \mu_2 \mu_3}_{\rm vac}>$ does not necessarily satisfy $<\partial_{\mu_0} \chi^{\mu_0 \mu_1 \mu_2 \mu_3}_{\rm vac}> =0$. So the total $J^{\mu_0 \mu_1 \mu_2 \mu_3}$ is not locally conserved at the quantum level, and the gauge transformation $\delta A_{\mu_0 \mu_1 \mu_2 \mu_3} = 4 \partial_{[\mu_0}\Lambda_{\mu_1 \mu_2 \mu_3 ]}$ generally induces a nonzero variation of $I_{\rm brane}^{(R)}$. Integrating by part one obtains from (3.31) that
\begin{equation}
\delta_G I_{\rm brane}^{(R)} = - \frac{1}{3!} \int d^4 x \Lambda_{\mu_1 \mu_2 \mu_3} <\partial_{\mu_0} \chi^{\mu_0 \mu_1 \mu_2 \mu_3}_{\rm vac}> \,\,,
\end{equation}
which generally takes nonzero values because so does  $<\partial_{\mu_0} \chi^{\mu_0 \mu_1 \mu_2 \mu_3}_{\rm vac}>$.

In addition to (3.35), there is another important variation of $I_{\rm brane}^{(R)}$ which plays a crucial role in our self-tuning mechanism. To find its explicit form, rewrite $\delta_G I_{\rm brane}^{(R)}$ as $\delta_G I_{\rm brane}^{(R)} = \mu_3 \int d \Lambda_{(3)}$ and take an ansatz \cite{1}
\begin{equation}
\Lambda_{(3)} = F(y) \sqrt{-g_4} dx^1 \wedge dx^2 \wedge dx^3 \,\,,
\end{equation}
where $F(y)$ is an arbitrary function of the internal coordinates $y^m$. (3.36) is the most appropriate ansatz which accords with (3.34) and therefore respects the Poincar$\acute{\rm e}$ symmetry of our four-dimensional spacetime. Once we take $\Lambda_{(3)}$ as in (3.36), $\delta_G I_{\rm brane}^{(R)}$ in (3.35) vanishes because $\partial_{[\mu_0} \Lambda_{\mu_1 \mu_2 \mu_3 ]} =0$ for a constant $\sqrt{-g_4}$. (Indeed $\sqrt{-g_4}$ is constant when $\lambda =0$. See below.) However, $\Lambda_{(3)}$ in (3.36) generates another type of $\delta_G I_{\rm brane}^{(R)}$ as shown below.

Taking derivative to $\Lambda_{(3)}$ one obtains
\begin{equation}
\delta_G I_{\rm brane}^{(R)} =  \int d^4 x \sqrt{-g_4} f_m (y) J^{m123} + \frac{3}{2} \int d^4 x \sqrt{-g_4} H\,F(y) J^{0123} \,\,,
\end{equation}
where $f_m (y)$($\equiv \partial_m F(y)$) represents $\delta A_{m123} / \sqrt{-g_4}$, and $H$($\equiv (2/3) \partial_0 \ln \sqrt{-g_4}$) is the Hubble constant of the four-dimensional spacetime $ds_4^2 = -dt^2 + e^{Ht} d {\vec x}_3$, which therefore vanishes for $\lambda=0$ because $\lambda \propto H^2$. In (3.37), $J^{m123}$ is defined by
\begin{equation}
J^{m123} = \mu_3 \epsilon^{\alpha_0 \alpha_1 \alpha_2 \alpha_3} \big(\frac{\partial Y^m}{\partial x^{\alpha_0}} \big) \wedge \big(\frac{\partial X^1}{\partial x^{\alpha_1}} \big) \wedge \big(\frac{\partial X^2}{\partial x^{\alpha_2}} \big) \wedge \big(\frac{\partial X^3}{\partial x^{\alpha_3}} \big)\,\,,
\end{equation}
which, at the classical level, vanishes for the embedding $X^{\mu}_{\rm cl} (x) = x^{\mu}$ because ${\partial Y^m_{\rm cl}}/{\partial x^{\alpha_0}}=0$. So the nonzero contribution to $J^{m123}$ comes from the quantum excitations $<\chi^{m123}_{\rm vac}>$. Denoting $<\chi^{m123}_{\rm vac}>$ by $\delta \mu_{\rm T}^m (\phi)$ (and omitting the second term) one can rewrite (3.37) as
\begin{equation}
\delta_G I_{\rm brane}^{(R)} =  \big(\int d^4 x \sqrt{-g_4}\big) \int dy \sqrt{h_6} \delta \mu_{\rm T}^m (\phi) f_m (y) \delta^6 (y) \,\,,
\end{equation}
where $f_m (y)$'s are arbitrary functions of $y^m$, representing (derivatives of) local gauge parameters.

\vskip 0.3cm
\hspace{-0.6cm}{\bf \large 3.5 Brane action density ${\hat I}_{\rm brane}$ and a new self-tuning mechanism}
\vskip 0.15cm

From (3.3) and (3.39) (or from (8.1)) one finds that at the quantum level the brane action consists of various parts,
\begin{equation}
I_{\rm brane} = \Big( I_{\rm brane}^{(NS)}({\rm tree}) + I_{\rm brane}^{(R)}({\rm tree}) \Big) + \Big( \delta_Q I_{\rm brane}^{(NS)} + \delta_Q I_{\rm brane}^{(R)} \Big) + \delta_G I_{\rm brane}^{(R)}\,\,.
\end{equation}
Among these terms, $I_{\rm brane}^{(NS)}(\rm tree)$ and $I_{\rm brane}^{(R)}(\rm tree)$ are the tree level actions and they always cancel out by field equations for the BPS $D3$-branes. (See, for instance, Sec VI.C of \cite{1} for this.) The correction terms $\delta_Q I_{\rm brane}^{(NS)}$ and $\delta_Q I_{\rm brane}^{(R)}$ arise from $\rho_{\rm vac} (\phi)$ and $\delta \mu (\phi)$, and they represent quantum fluctuations (of the gravitational and standard model fields with support) on the $D3$-brane. So $\delta_Q I_{\rm brane}^{(NS)} + \delta_Q I_{\rm brane}^{(R)}$ correspond to the gravitational plus electroweak and QCD vacuum energies of the standard model configurations of the brane region. These two terms are conjectured to cancel out to all orders of perturbations in supersymmetric theories, but they do not when supersymmtry (of the brane region) is broken. (The cancelation at one-loop order on the BPS $D3$-branes has been proven explicitly for the case $G_{(3)} =0$ in Sec. VIIA of \cite{1}.)  In our self-tuning mechanism, however, it is not important whether such a cancelation occurs or not, as we will see in what follows.

Using $\beta = 4 \lambda$, one can rewrite (3.30) as
\begin{equation}
\lambda = \frac{1}{24} \chi^{1/2} ({\mathcal N} -1)({\mathcal N} -3)\big(1- 3 b_0 \Pi({\mathcal N})\big)V \,\,.
\end{equation}
(3.41) requires that $\lambda$ must vanish once $V$ belongs to $V_n$ ($V \in V_n$) with $n=1$ or 3, where $V_n$ represents a class of potential densities satisfying
\begin{equation}
{\mathcal N} V_n = n V_n \,\,.
\end{equation}
Aside from this, one also finds that if $V \in V_n$, (3.20) becomes
\begin{equation}
\lambda = \frac{(n-1)}{4} \kappa^2 {\mathcal V}_{\rm scalar} + \frac{\kappa^2}{2} {\hat I}_{\rm brane} \,\,
\end{equation}
by (3.42) and (3.17). So if $V \in V_1$, $\lambda$ is simply $\lambda = \frac{\kappa^2}{2} {\hat I}_{\rm brane}$. But if $V \in V_3$, then $\lambda$ becomes
\begin{equation}
\lambda = \frac{\kappa^2}{2} \Big({\mathcal V}_{\rm scalar} + {\hat I}_{\rm brane} \Big) \,\,,
\end{equation}
and in both cases $\lambda$ vanishes by (3.41). In our scenario proposed in Sec. 5.2, the background vacuum of our present universe is identified with one of the AdS vacua of KKLT, and in Secs. IV and VI it will be shown that these AdS vacua all belong to $V_3$, $V_{\rm AdS} \in V_3$. So in our case $\lambda$ is basically given by (3.44) and it must vanish by the self-tuning equation (3.41). (But in the next paragraphs we will show that (3.44) becomes (1.1) by (3.40). So in our AdS vacuum scenario $\lambda$ is basically given by (1.1) and it must vanish by the self-tuning equation (3.41).)

Let us go back to (3.40). We have seen in Sec. 3.4 that the last term $\delta_G {\hat I}_{\rm brane}^{(R)}$ represents the magnitude of gauge symmetry breaking of ${\hat I}_{\rm brane}^{(R)}$ (caused by an anomaly $<\partial_m \chi_{\rm vac}^{m123}> \neq 0$) arising at the quantum level, where $<\chi_{\rm vac}^{m123}>$ are quantum excitations on the branes with components along the transverse directions of the $D3$-branes. In \cite{1}, it was shown that $\delta_G {\hat I}_{\rm brane}^{(R)}$ is closely related to the supersymmetry breaking of the brane region. It plays the role of a supersymmetry breaking term. (The supersymmetry breaking caused by $\delta_G {\hat I}_{\rm brane}^{(R)}$ is also discussed in detail in Sec. VIII of this paper.) So $\delta_G {\hat I}_{\rm brane}^{(R)}$ is an energy scale of the gauge symmetry breaking (or an anomaly) of the action ${\hat I}_{\rm brane}^{(R)}$, and at the same time it is also an energy scale of the supersymmetry breaking induced by this gauge symmetry breaking of ${\hat I}_{\rm brane}^{(R)}$.

After all, renaming $\delta_G {\hat I}_{\rm brane}^{(R)}$ as
\begin{equation}
\delta_G {\hat I}_{\rm brane}^{(R)} \equiv -{\mathcal E}_{\rm SB} \,\,,
\end{equation}
one obtains (1.1) from (3.40) and (3.44) (Recall that the tree level actions ${\hat I}_{\rm brane}^{(NS)} (\rm tree) + {\hat I}_{\rm brane}^{(\it R)} (\rm tree)$ cancel out for the BPS $D3$-branes.), where ${\mathcal E}_{\rm SB}$ is now
\begin{equation}
{\mathcal E}_{\rm SB} = - \int d^6 y \sqrt{h_6} \delta \mu_{\rm T}^{m} (\phi) f_m (y) \delta^6 (y) \,\,
\end{equation}
from (3.39) and (3.45). Note that ${\mathcal E}_{\rm SB}$ contains arbitrary gauge parameters $f_m (y)$. This implies that ${\mathcal E}_{\rm SB}$ possesses gauge arbitrariness. Due to this property of ${\mathcal E}_{\rm SB}$, any nonzero values of ${\mathcal V}_{\rm scalar}$ and $\delta_Q {\hat I}_{\rm brane}^{(NS)} + \delta_Q {\hat I}_{\rm brane}^{(R)}$ in (1.1) can be gauged away by this ${\mathcal E}_{\rm SB}$, and as a result $\lambda$ vanishes (by (3.41)) as long as the potential density $V$ satisfies $V \in V_n$ with $n=1$ or 3.

The above self-tuning mechanism is distinguished from the usual theories where $\lambda$ is directly identified with ${\mathcal V}_{\rm scalar}$. In those theories, $\lambda$ is generally unstable under perturbative (radiative) corrections because so is ${\mathcal V}_{\rm scalar}$. Also the dS vacua necessarily imply ${\mathcal V}_{\rm scalar} > 0$, which can lead to a tunneling instability as mentioned in the opening paragraph of this section. But in our self-tuning mechanism described above, these are not to be the cases anymore. $\lambda$ can vanish by (3.41) regardless of whether ${\mathcal V}_{\rm scalar}$ in (1.1) vanishes or not. So we can take ${\mathcal V}_{\rm scalar} < 0$ (while maintaining $\lambda =0$) to avoid the tunneling instability, and the value $\lambda =0$ is always stable against radiative corrections. Any nonzero contributions to ${\mathcal V}_{\rm scalar}$ and quantum fluctuations on the branes are forced to be gauged away by (3.41) as long as $V$ satisfies $V \in V_n$ with $n=1$ or 3, and $\lambda =0$ is automatically achieved in our self-tuning mechanism of this paper. Hence in the following sections we will mainly check if our background configurations really satisfy $V \in V_n$ with $n=1$ or 3.

\vskip 1cm
\hspace{-0.65cm}{\bf \Large IV. AdS vacua of KKLT and gravitino mass}
\vskip 0.5cm
\setcounter{equation}{0}
\renewcommand{\theequation}{4.\arabic{equation}}

In no-scale structure (and in the ISD background) $\lambda$ trivially vanishes from (3.41) because potential densities arising from ISD fluxes all vanish. But once the no-scale structure is broken by nonperturbative effects as in AdS vacua of KKLT, the potential density does not vanish anymore because $G_{(3)}$ now acquires IASD components due to the presence of nonperturbative terms in the superpotential $W$. Besides this, the scalar potential (2.4) receives nontrivial contributions from both perturbative and nonperturbative corrections of the superpotential and the K${\ddot{\rm a}}$hler potential. More explicitly, while the superpotential receives only nonperturbative corrections $W=W_{\rm tree} +W_{\rm np}$ as in (2.1) \cite{11}, the K${\ddot{\rm a}}$hler potential receives both perturbative and nonperturbative corrections ${\mathcal K}={\mathcal K}_{\rm tree} + {\mathcal K}_{\rm p} +{\mathcal K}_{\rm np}$. So in order to maintain $\lambda=0$, the potential density of our background vacuum must remain to satisfy (3.42) with $n=1,3$ even under these corrections. In this section we want to show that the AdS vacua of KKLT satisfy the above property. Namely the potential densities of AdS vacua of KKLT belong to $V_3$, and this result does not change under $W=W_{\rm tree} +W_{\rm np}$ and ${\mathcal K}={\mathcal K}_{\rm tree} + {\mathcal K}_{\rm p} +{\mathcal K}_{\rm np}$.

\vskip 0.3cm
\hspace{-0.6cm}{\bf \large 4.1 AdS vacua of KKLT}
\vskip 0.15cm

\vskip 0.3cm
\hspace{-0.6cm}{(1) Scalar potential of the AdS vacua}
\vskip 0.15cm

The scalar potential arising from the fluxes can be obtained from the $G_{mnp}{\bar G}^{mnp}$ term of the action (see (3.6)). Rewrite the $G_{mnp}{\bar G}^{mnp}$ term as\footnote{In obtaining (4.1) we have used the identity $G_{(3)} \wedge \ast_6 {\bar G}_{(3)}= - i G_{(3)} \wedge {\bar G}_{(3)} + 2i G_{(3)}^+ \wedge {\bar G}_{(3)}^{+}$.} \cite{6}
\begin{equation}
-\frac{1}{24 \kappa_{10}^2} \int d^6 y \sqrt{h_6} \,e^{2B} G_{mnp}{\bar G}^{mnp} =\frac{i}{4 \kappa_{10}^2 Im \tau} \int \frac{\chi^{1/2}}{g_s} G_{(3)} \wedge {\bar G}_{(3)} ~~~~~~~~~~~~~~~~~~~~\nonumber
\end{equation}
\begin{equation}
~~~~~~~~~~- \frac{1}{12 \kappa_{10}^2 Im \tau} \int d^6 y \sqrt{h_6} \, \frac{\chi^{1/2}}{g_s} G_{mnp}^{+}{\bar G}^{+mnp} \,\,,
\end{equation}
where
\begin{equation}
G_{(3)}^{\pm} = \frac{1}{2} ( G_{(3)} \pm i \ast_6 G_{(3)})\,\,, ~~~~~~ \ast_6 G_{(3)}^{\pm} =\mp i G_{(3)}^{\pm} \,\,,
\end{equation}
are the IASD/ISD parts of $G_{(3)}$, $G_{(3)}^+ = G_{(3)}^{\rm IASD}$ and $G_{(3)}^- = G_{(3)}^{\rm ISD}$. The scalar potential (arising from the fluxes) is defined by the second term of (4.1) as
\begin{equation}
{\mathcal V}_{\rm no-scale} =  \frac{1}{12 \kappa_{10}^2 Im \tau} \int d^6 y \sqrt{h_6} \, \frac{\chi^{1/2}}{g_s} G_{mnp}^{+}{\bar G}^{+mnp} \,\,,
\end{equation}
which is identified with (2.5) of the four-dimensional effective theory (see Sec. A.2 of  \cite{6}). Since ${\mathcal V}_{\rm no-scale}$ is given as to be $\propto \int G_{(3)}^{\rm IASD} \cdot {\bar G}_{(3)}^{\rm IASD}$, it (and its density as well) vanishes in the ISD compactifications where the superpotential is simply given by (2.6). But once the nonperturbative term is added as in (2.1), $G_{(3)}$ can not remain ISD anymore. The unbroken supersymmetry $DW =0$ requires that $G_{(3)}$ must also contain (1,2) and (3,0) components (see for instance \cite{9}) in addition to the ISD components. Hence in this case (4.3) receives nonzero contributions from these fluxes.

The nonzero density of (4.3), however, satisfies $V_{\rm no-scale} \in V_3$, so it does not contribute to $\lambda$ in (3.41). But (4.3) is only referred to the no-scale type potential (2.5). In the AdS vacua of KKLT there is another important contribution to ${\mathcal V}_{\rm scalar}$ coming from the nonperturbative superpotential (2.1). Namely from (2.7) one obtains
\begin{equation}
{\mathcal V}_{\rm AdS} = - \frac{3}{2 \kappa_{10}^2} e^{\mathcal K} |W|^2 \,\,
\end{equation}
under $D_\rho W=0$, which reduces to (2.8) by (2.1). (4.4) includes nonperturbative correction because $W$ in (4.4) contains the term $A e^{-a \sigma}$. The nonperturbative term can arise for instance from the gaugino condensation on $D7$-branes wrapping a four-cycle of the internal space \cite{5}. In the heterotic string theory the three-form structure of the potential density with a gaugino condensation $<{\rm tr}{\bar \lambda}\Gamma^{mnp}\lambda>$ is manifest in the action \cite{12},
\begin{equation}
I_{\rm het}= - \frac{1}{2 \kappa_{10}^2} \int e^{-\phi} \big( H_{(3)} - \frac{\alpha^{\prime}}{16} e^{\phi/2} {\rm tr}{\bar \lambda}\Gamma_{(3)}\lambda \big)^2 \,\,.
\end{equation}
So the potential density associated with (4.5) obviously belongs to $V_3$. In the case of type IIB theory, however, the tensor structure of (4.4) is not quite obvious at this point and we need some procedure to find it out.

\vskip 0.3cm
\hspace{-0.6cm}{(2) $ \int  \Omega \wedge {\bar \Omega}$ structure of the superpotential $W$}
\vskip 0.15cm

Since $G_{(3)}$ generally contains both ISD and IASD components, we decompose $G_{(3)}$ as
\begin{equation}
G_{(3)} = \alpha_0 \Omega + {\bar \alpha}_0 {\bar \Omega} + \beta^I \chi_I +  {\bar \beta}^{\bar I} {\bar \chi}_{\bar I} \,\,,
\end{equation}
where $\Omega$ is the holomorphic (3,0)-form and $\chi_I$ denotes the basis of $H^{(2,1)}$. Then using (4.6) one can express $W_0$ in (2.6) as
\begin{equation}
W_0 = - {\bar \alpha}_0 \int \Omega \wedge {\bar \Omega} \,\,.
\end{equation}
Apart from this, on the other hand, the total superpotential (2.1) satisfies
\begin{equation}
D_I W=0
\end{equation}
at the supersymmetric (AdS) minimum $\sigma = \sigma_m$ of KKLT, where the index $I$ labels the complex structure moduli. From (2.3) and the definition of the covariant derivative $D_I$ one finds that (4.8) requires $W$ to take the form
\begin{equation}
W= -{\bar \alpha} \int  \Omega \wedge {\bar \Omega} \,\,,
\end{equation}
where ${\bar \alpha}$ can depend on $\tau$, but not on the complex structure moduli. The above result suggests that the $\int \Omega \wedge \bar \Omega$  structure of the superpotential $W_0$ remains unchanged under nonperturbative corrections. To understand this more explicitly, we consider the following discussions.

\vskip 0.3cm
\hspace{-0.6cm}{(3) Nonperturbative corrections and deformation of the complex structure}
\vskip 0.15cm

When the nonperturbative term $A e^{-a \sigma}$ is absent the complex structure moduli of the superpotential $\int_{{\mathcal M}_6} G_{(3)} \wedge \Omega(\tau^I )$ ($\equiv W(\tau^I )$) are stabilized at values $\tau^I = \tau_0^I$ for which $G_{(3)}$ is ISD at the tree level. Hence in this case the superpotential of the stabilized minimum is just $W(\tau_0^I )$, which is identified with $W_0$ in (2.6). Let us now introduce the nonperturbative term $A e^{-a \sigma}$. Once we introduce this term, the point $\tau^I = \tau_0^I$ in the moduli space would not be the stabilized point anymore. It deviates from $\tau^I = \tau_0^I$ along the complex structure moduli direction by the same amount of the nonperturbative corrections. So the new stabilized point becomes $\tau^I = \tau_0^I + \delta \tau^I$ and the superpotential of the supersymmetric minimum also changes from $W(\tau_0^I )$ to $W(\tau_0^I + \delta \tau^I )$, where $W(\tau_0^I + \delta \tau^I )$ now includes the nonperturbative correction and the deviation $\delta \tau^I$ will be determined by the nonperturbative term because the former is generated by the latter. To see this more explicitly, in the followings we will decompose $W(\tau_0^I + \delta \tau^I )$ into $W_0 + \delta W(\tau_0^I )$ and identify $\delta W (\tau_0^I )$ with the nonperturbative term $Ae^{-a \sigma_m}$.

\vskip 0.3cm
\hspace{-0.6cm}{(4) Decomposition of $W(\tau_0^I + \delta \tau^I )$}
\vskip 0.15cm

Since $W(\tau^I )$ is given by $\int_{{\mathcal M}_6} G_{(3)} \wedge \Omega(\tau^I )$, we have
\begin{equation}
W(\tau_0^I + \delta \tau^I ) = \int_{{\mathcal M}_6} G_{(3)}^{\rm NEW} \wedge \Omega (\tau_0^I ) + \int_{{\mathcal M}_6} G_{(3)}^{\rm NEW} \wedge \delta \Omega (\tau_0^I )\,\,,
\end{equation}
where
\begin{equation}
\delta \Omega (\tau_0^I )= \delta \tau^J \partial_J \Omega (\tau_0^I ) + O(({\delta \tau^J})^2 ) \,\,.
\end{equation}
In (4.10), $G_{(3)}$ has been replaced by the new three-form $G_{(3)}^{\rm NEW}$ ($\equiv G_{(3)} + \delta G_{(3)}$) because the stabilized point has been changed into a new one by the nonperturbative correction, and the complex structure moduli of the new stabilized point cannot be fixed by the original $G_{(3)}$ in $W_0$. So in this section we temporarily use $G_{(3)}^{\rm NEW}$ to denote the three-form fluxes in $W(\tau^I )$, to distinguish it from the original $G_{(3)}$ in $W_0$. Now the first term of (4.10) can be decomposed into $W_0$ plus $\int_{{\mathcal M}_6} \delta G_{(3)} \wedge \Omega(\tau_0^I )$ because $\int_{{\mathcal M}_6} G_{(3)} \wedge \Omega(\tau_0^I )$ is just $W_0$ as mentioned in the subsection (3). Similarly, the second term of (4.10) can be also decomposed into $\int_{{\mathcal M}_6} G_{(3)} \wedge \delta \Omega (\tau_0^I )$ plus $\int_{{\mathcal M}_6} \delta G_{(3)} \wedge \delta \Omega (\tau_0^I )$, but where $\int_{{\mathcal M}_6} \delta G_{(3)} \wedge \delta \Omega (\tau_0^I )$ can be neglected in the leading order approximation because it is already of the second order in $\delta \tau^I$.

In (4.11) $\partial_J \Omega (\tau_0^I )$ represent the values of $\partial_J \Omega (\tau^I )$ at $\tau^I = \tau_0^I$, and if we use the well-known formula $\partial_J \Omega = (- \partial_J {\mathcal K}) \Omega + \chi_J$, one can rewrite (4.11) as
\begin{equation}
\delta \Omega (\tau_0^I )= (-\delta {\mathcal K}) \Omega (\tau_0^I ) + \delta \tau^J \chi_J (\tau_0^I) + O(({\delta \tau^J})^2 ) \,\,,
\end{equation}
where $\delta {\mathcal K}$ ($\equiv \delta \tau^J \partial_J {\mathcal K}$) is the variation of the K$\ddot{\rm a}$hler potential ${\mathcal K}$ caused by the nonperturbative correction $A e^{-a \sigma_m}$. Using (4.12) one can rewrite (4.10) as
\begin{equation}
W(\tau_0^I + \delta \tau^I )= W_0 +\delta W(\tau_0^I ) \,\,,
\end{equation}
where $\delta W(\tau_0^I )$ is now given by
\begin{equation}
\delta W(\tau_0^I ) =  \int_{{\mathcal M}_6} \delta G_{(3)} \wedge \Omega (\tau_0^I ) +(-\delta {\mathcal K}) \int_{{\mathcal M}_6} G_{(3)} \wedge \Omega (\tau_0^I ) + {\delta \tau^J} \int_{{\mathcal M}_6} G_{(3)} \wedge \chi_J (\tau_0^I )+ O(({\delta \tau^J})^2 ) \,\,.
\end{equation}

\vskip 0.3cm
\hspace{-0.6cm}{(5) $\int  \Omega \wedge {\bar \Omega}$ structure of the superpotential $W$ again}
\vskip 0.15cm

Turning back to the nonperturbative term $A(\tau^I )e^{-a \sigma_m}$, let us consider the properties of $A(\tau^I)$. In ten-dimensional pictures the complex structure moduli of the Calabi-Yau threefolds are encoded in the harmonic three-form basis $\Omega$ and $\chi_I$. So in the ten-dimensional picture the scalar functions of the complex structure moduli, such as $A(\tau^I)$, in four-dimensional effective theory must appear essentially in terms of (or as linear combinations of) the nonzero six-dimensional integrals $\int \Omega \wedge \bar \Omega$ and $\int \chi_I \wedge {\bar \chi}_{\bar J}$ as in (4.14) because the four-dimensional effective theory is obtained by a dimensional reduction of the ten-dimensional theory.

Now we identify $\delta W(\tau_0^I )$ in (4.14) with $A(\tau^I )e^{-a \sigma_m}$ as mentioned in the subsection (3) :
\begin{equation}
 \int_{{\mathcal M}_6} \delta G_{(3)} \wedge \Omega (\tau_0^I )+ (-\delta {\mathcal K}) \int_{{\mathcal M}_6} G_{(3)} \wedge \Omega (\tau_0^I )+ {\delta \tau^J} \int_{{\mathcal M}_6} G_{(3)} \wedge \chi_J (\tau_0^I ) + O(({\delta \tau^J})^2 ) \nonumber
\end{equation}
\begin{equation}
\equiv A(\tau_0^I + \delta \tau^I )e^{-a \sigma_m} \,\,.
\end{equation}
The above identification may be achieved by adjusting ${\delta \tau^J}$ and $\delta G_{(3)}$ properly, which means that ${\delta \tau^J}$ and $\delta G_{(3)}$ are entirely determined by the nonperturbative term $A e^{-a \sigma_m}$. As an example, we take  ${\delta \tau^J} = e^{-a \sigma_m} {\delta \tilde{\tau}^J}$, ${\delta G_{(3)}} = e^{-a \sigma_m} {\delta \tilde{G}_{(3)}}$, and adjust ${\delta \tilde{\tau}^J}$ and ${\delta \tilde{G}_{(3)}}$ properly so that (4.15) is satisfied. Then in the leading order approximation ${\delta \tilde{\tau}^J}$ and ${\delta \tilde{G}_{(3)}}$ are determined from the coefficients of the Hodge decompositions $\int \Omega \wedge \bar{\Omega}$ and $\int \chi_I \wedge \bar{\chi}_{\bar J}$ of $A(\tau_0^I )$. In this way, from (4.13) and (4.15) we finally have
\begin{equation}
W(\tau_0^I + \delta \tau^I ) = W_0 + A(\tau^I )e^{-a \sigma_m} \,\,,
\end{equation}
where $W(\tau_0^I + \delta \tau^I )$ is given by $\int_{{\mathcal M}_6} G_{(3)}^{\rm NEW} \wedge  \Omega (\tau_0^I + \delta \tau^I )$ (see (4.10)).

The above result shows that the combined (total) superpotential $W_0 + A e^{-a \sigma_m}$ in (2.1) can be written in the form
\begin{equation}
W (\tau^I ) = \int_{{\mathcal M}_6} G_{(3)}^{\rm NEW} \wedge  \Omega (\tau^I ) \,\,,
\end{equation}
where $\tau^I$ denotes $\tau_0^I + \delta \tau^I$. $W(\tau^I )$ in (4.17) has the $\int \Omega \wedge \bar \Omega$ structure just like $W_0$ in (2.6) despite that $A e^{-a \sigma_m}$ in $W(\tau^I )$ includes both $(3,0)\bigotimes(0,3)$ and $(2,1)\bigotimes(1,2)$ terms (see (4.15)). Indeed the last term ${\delta \tau^J} \int_{{\mathcal M}_6} G_{(3)} \wedge \chi_J (\tau_0^I )$ of (4.15) vanishes if $W_0$ satisfies $D_I W_0 =0$. But the F-term condition of the stabilized point has now been changed into $D_I W (\tau^I )=0$ from $D_I W_0 =0$. But still if we neglect the higher order terms, the last term of (4.15) can be neglected because $D_I W (\tau^I )=0$ approximately requires that $\delta \tau^J \int_{{\mathcal M}_6} G_{(3)} \wedge \chi_{J} (\tau_0^I )$ must vanish. So in this approximation the total superpotential $W (\tau^I )$ can be written as
\begin{equation}
W (\tau^I ) = \int_{{\mathcal M}_6} G_{(3)}^{\rm EFF} \wedge  \Omega (\tau_0^I )\,\,,
\end{equation}
where $G_{(3)}^{\rm EFF}$ is defined by $G_{(3)}^{\rm EFF} \equiv (1- \delta {\mathcal K}) G_{(3)} + \delta G_{(3)}$.

The above $W (\tau^I )$ is of the same form as $W_0$ in (2.6) only except that $G_{(3)}$ in (2.6) is replaced by $G_{(3)}^{\rm EFF}$. According to (4.18), $W(\tau^I )$ may be regarded as a superpotential generated by an effective three-form flux $G_{(3)}^{\rm EFF}$, where the nonperturbative effects are merged with $G_{(3)}$ to form $G_{(3)}^{\rm EFF}$. So at least at the supersymmetric minimum the effect of the nonperturbative term $A e^{-a \sigma_m}$ in $W(\tau^I )$ is to change $G_{(3)}$ into a new flux $G_{(3)}^{\rm EFF}$ which also has the three-form structure like the original $G_{(3)}$.

\vskip 0.3cm
\hspace{-0.6cm}{(6) Tensor structure of ${\mathcal V}_{\rm AdS}$}
\vskip 0.15cm

We have just seen that the total superpotential $W(\tau^I )$ has the $(3,0)\bigotimes(0,3)$ structure in the leading order approximation where the terms of order higher than $(\delta \tau^{I})^2$ are neglected. In our case, however, we don't have to use this approximation to find the Hodge structure of $W (\tau^I )$. We already know from (4.17) that the superpotential $W(\tau^I )$ has the $\propto \int \Omega \wedge {\bar \Omega}$ structure just like $W_0$. The only difference between $W_0$ and $W(\tau^I )$ is that $\Omega (\tau_0^I )$ and $G_{(3)}$ in $W_0$ are now replaced by $\Omega (\tau^I )$ and $G_{(3)}^{\rm NEW}$ in (4.17), and hence also in (4.7) and (4.9), $\Omega$ in $W$ is $\Omega (\tau^I )$, while $\Omega$ in $W_0$ is $\Omega (\tau_0^I )$. (The fact that $\Omega$ in (4.9) is $\Omega (\tau^I )$ means that the holomorphic three-forms contained in (2.3) also change from $\Omega (\tau_0^I )$ to $\Omega (\tau^I )$ under the nonperturbative correction because (4.9) is obtained from (2.3) by $D_I W=0$.) Despite these differences, however, the Hodge structures of $W$ and $W_0$ are entirely identical. They are both $\propto \int \Omega \wedge {\bar \Omega}$ (or $\int_{{\mathcal M}_6} G_{(3)} \wedge  \Omega$), though the complex structures of each $\Omega$'s in $W$ and $W_0$ are different from one another.

Now using (4.17) one can determine the tensor structure of ${\mathcal V}_{\rm AdS}$ in (4.4). (4.17) shows that $W$ acquires nonzero values from the $(0,3)$ component of $G_{(3)}^{\rm NEW}$ ($\equiv G_{(0,3)}^{\rm NEW}$). Writing $G_{(0,3)}^{\rm NEW}$ as $G_{(0,3)}^{\rm NEW} = {\bar \alpha}{\bar \Omega}$, one obtains
\begin{equation}
\int {\bar G}_{(0,3)}^{\rm NEW} \wedge G_{(0,3)}^{\rm NEW} = - \frac{\int G_{(0,3)}^{\rm NEW} \wedge \Omega \int {\bar G}_{(0,3)}^{\rm NEW} \wedge {\bar \Omega}}{\int \Omega \wedge {\bar \Omega}} \,\,,
\end{equation}
and using $\ast_6 G_{(0,3)}^{\rm NEW}= i G_{(0,3)}^{\rm NEW}$ (note that $G_{(0,3)}^{\rm NEW}$ is ISD) one finds that ${\mathcal V}_{\rm AdS}$ in (4.4) becomes
\begin{equation}
{\mathcal V}_{\rm AdS} =  \frac{1}{4 \kappa_{10}^2} \, e^{\mathcal K_{\tau} + {\mathcal K}_{\rho} } \int d^6 y \sqrt{h_6} \big( G_{(0,3)}^{\rm NEW} \big)_{mnp} \big({\bar G}_{(0,3)}^{\rm NEW}\big)^{mnp}\,\,.
\end{equation}
(4.20) shows that the density of ${\mathcal V}_{\rm AdS}$ clearly belongs to $V_3$.

In (4.20) ${\mathcal V}_{\rm AdS}$ receives nonzero contribution from the ISD $(0,3)$ component of $G_{(3)}^{\rm NEW}$ as opposed to the case of (4.3) where  ${\mathcal V}_{\rm no-scale}$ receives nonzero contributions only from the IASD components of $G_{(3)}$. Indeed $G_{(3)}^{\rm NEW}$ in (4.17) contains only $(3,0)$ component as an IASD piece because (4.8) requires that $(1,2)$ component of $G_{(3)}^{\rm NEW}$ should vanish. In any case, both ${\mathcal V}_{\rm no-scale}$ and ${\mathcal V}_{\rm AdS}$ in (4.3) and (4.20) take nonzero values in the AdS vacua of KKLT even at the tree level. However, they never contribute to $\lambda$ in (3.41) because their densities $V_{\rm no-scale}$ and $V_{\rm AdS}$ both belong to $V_3$. Hence $\lambda$ must be self-tuned to vanish in the AdS vacua of KKLT.

\vskip 0.3cm
\hspace{-0.6cm}{\bf \large 4.2 Gravitino mass}
\vskip 0.15cm

The scalar potential arising from (2.1) does not vanish at the supersymmetric minimum of the potential. In general it is proportional to $|W_0 |^2$, or more precisely,
\begin{equation}
{\mathcal V}_{\rm scalar} \propto  e^{\mathcal K} |W_0 |^2 \,\,,
\end{equation}
at the extremum of the potential \cite{13,14}. Indeed at the AdS minimum $D_{\rho} W=0$ of KKLT, the coefficient $A$ is given by
\begin{equation}
A = - W_0 \, e^{a \sigma_{m}} \big(1+\frac{2}{3}a \sigma_{m} \big)^{-1} \,\,,
\end{equation}
and therefore the scalar potential (2.8) becomes proportional to $e^{\mathcal K} |W_0 |^2$ there,
\begin{equation}
{\mathcal V}_{\rm AdS} = \frac{1}{2 \kappa_{10}^2} \Big( -\frac{a^2}{6 \sigma_m} \big( 1+ \frac{2}{3} a \sigma_{m}\big)^{-2} \,\Big)\, e^{\mathcal K_{\tau} + {\mathcal K}_{\rm cs} } |W_0 |^2 \,\,,
\end{equation}
where ${\mathcal K}_{\rm cs}$, $\mathcal K_{\tau}$ are the K${\ddot{\rm a}}$hler potentials for the complex structure moduli and the axion/dilaton, respectively.

(4.22) shows that $W_0$ necessarily takes nonzero values in the presence of the nonperturbative correction $A e^{-a \sigma_{m}}$. The nonzeroness of $W_0$ implies that $G_{(3)}$ must contain $(0,3)$ component, and in the presence of this component the gravitino generally acquires nonzero mass $m_{3/2}$ from the $G_{(3)}$ flux. The gravitino mass term of the reduced action for the type IIB theory can be obtained through the decomposition
\begin{equation}
\Psi_\mu = \psi_\mu \otimes e^{\frac{B}{4}} \eta \,\,,
\end{equation}
where $\Psi_\mu / \psi_\mu$ are the ten$/$four-dimensional gravitini, respectively, and $\eta$ is a six-dimensional killing spinor satisfying $\gamma^{\bar{i}} \eta =0$, where $\gamma^{\bar{i}}$ is the six-dimensional Dirac matrix represented in the complex basis. In the real basis of the Calabi-Yau one obtains \cite{14}
\begin{equation}
I_{3/2} = \frac{1}{\kappa_{10}^2} \int d^4 x \sqrt{-g_4} \frac{1}{(Im \rho)^{3/2}}\, \Big\{ \big(\bar{\psi}_\mu \gamma^{\mu \nu} \psi_{\nu}^{\ast} \big) \big(\frac{i}{48} \int d^6 y \sqrt{h_6}  \frac{1}{(Im \tau)^{1/2}} \eta^{+} \gamma^{mnp} \eta^{\ast} G_{mnp} \big) \nonumber
\end{equation}
\begin{equation}
+\, {\rm hermitian \,~conjugate\,~ term} \Big\}\,\,,
\end{equation}
where $m_{3/2}$ is identified as
\begin{equation}
m_{3/2} = \frac{\kappa^2}{\kappa_{10}^2}  \frac{1}{(Im \rho)^{3/2}} \frac{1}{(Im \tau)^{1/2}} \Big(\frac{1}{24} \int d^6 y \sqrt{h_6} \,\eta^{+} \gamma^{mnp} \eta^{\ast} G_{mnp} \Big)\,\,.
\end{equation}
Since all components except $\eta^{+} \gamma^{\bar{i}\bar{j}\bar{k}} \eta^{\ast}$ ($= \Omega^{\bar{i}\bar{j}\bar{k}}/ \|\Omega \|$) of $\eta^{+} \gamma^{mnp} \eta^{\ast}$ vanish by $\gamma^{\bar{i}} \eta =0$ in the complex basis, only the $(0,3)$ piece of $G_{(3)}$ contributes to $m_{3/2}$.

(4.25) shows that the density $V_{3/2}$ of $I_{3/2}$ is proportional to $\eta^{+} \gamma^{mnp} \eta^{\ast} G_{mnp}$. So $V_{3/2} \in V_3$, and the gravitino mass term $I_{3/2}$ arising from $G_{(0,3)}$ does not contribute to $\lambda$ just like ${\mathcal V}_{\rm AdS}$ of KKLT. Indeed, the gravitino mass $m_{3/2}$ is closely related to the ${\mathcal V}_{\rm AdS}$ of KKLT. Since $m_{3/2}^2$ is identified with $<e^{\mathcal K} |W_{0}|^2 >$ (see \cite{11} or \cite{14}), and $<e^{\mathcal K} |W_{0}|^2 >$ is a constant times ${\mathcal V}_{\rm AdS}$ at the AdS minimum (see (4.21) or (4.23)), $m_{3/2}^2$ is proportional to ${\mathcal V}_{\rm AdS}$ of KKLT. So one of the ways of ascertaining whether ${V}_{\rm AdS} \in V_3$ is really true is to check whether $m_{3/2}$ in (4.26) satisfies ${\hat m}_{3/2}^2 \in V_3$ or not, where ${\hat m}_{3/2}^2$ is the density of $m_{3/2}^2$ defined by $m_{3/2}^2 \equiv (1/2 \kappa_{10}^{2} g_s^2 \,) \int d^6 y \sqrt{h_6} \,{\hat m}_{3/2}^2$. The fact that ${\hat m}_{3/2}^2$ satisfies ${\hat m}_{3/2}^2 \in V_3$ can be proved easily as follows. Using $\ast_6 G_{(0,3)} = i G_{(0,3)}$ one can show that the square of (4.26), $m_{3/2}^2$, is proportional to $\big| \int G_{(0,3)} \wedge {\hat \Omega} \big|^2$, where ${\hat \Omega} \equiv \Omega / \|\Omega\|$. Next, using (4.19) one can show that $\big| \int G_{(0,3)} \wedge {\hat \Omega} \big|^2$ is proportional to $\int {\bar G}_{(0,3)} \wedge G_{(0,3)}$. Finally, using $\ast_6 G_{(0,3)} = i G_{(0,3)}$ again one can show that $\int {\bar G}_{(0,3)} \wedge G_{(0,3)}$ is proportional to $\int d^6 y \sqrt{h_6} (G_{(0,3)})_{mnp} ({\bar G}_{(0,3)})^{mnp}$. After all, one finds that ${\hat m}_{3/2}^2 \in V_3$ because $m_{3/2}^2$ is proportional to $\int d^6 y \sqrt{h_6} (G_{(0,3)})_{mnp} ({\bar G}_{(0,3)})^{mnp}$, which confirms the result of Sec. 4.1 that $V_{\rm AdS} \in V_3$.

The result $V_{\rm AdS} \in V_3$ is not affected by the perturbative and nonperturbative corrections ${\mathcal K} = {\mathcal K}_{\rm tree} + {\mathcal K}_{\rm p} +{\mathcal K}_{\rm np}$ and $W=W_{\rm tree} + W_{\rm np}$. The corrections ${\mathcal K}_{\rm p}$, ${\mathcal K}_{\rm np}$ in ${\mathcal K}$ act only as multiplicative factors $e^{{\mathcal K}_{\rm p}}$, $e^{{\mathcal K}_{\rm np}}$ in (4.20) (or in (4.21)), and on the other hand $W_{\rm np}$ in $W$ has already been considered in our discussions (namely in (2.1) and (6.3)). So the structure (4.20) of ${\mathcal V}_{\rm AdS}$, and consequently the result $V_{\rm AdS} \in V_3$ does not change by ${\mathcal K} = {\mathcal K}_{\rm tree} + {\mathcal K}_{\rm p}+ {\mathcal K}_{\rm np}$ and $W=W_{\rm tree} + W_{\rm np}$.

\vskip 1cm
\hspace{-0.65cm}{\bf \Large V. dS vacua of KKLT and AdS vacuum scenario}
\vskip 0.5cm
\setcounter{equation}{0}
\renewcommand{\theequation}{5.\arabic{equation}}

The next step of KKLT is to introduce $\overline{D3}$-branes (anti-$D3$-branes) at the end of the KS throat to obtain dS vacua. Introduction of $\overline{D3}$-branes induces an additional term $\delta {\mathcal V}_{\rm scalar}$ (see (2.9)) to the scalar potential as anticipated from the analysis of \cite{15}. Thus the scalar potential after introducing $\overline{D3}$-branes must be the sum of (2.8) and (2.9), where ${\mathcal V}_{\rm scalar}$ in (2.8), which is ${\mathcal V}_{\rm AdS}$ in fact, has already been verified to respects (3.42) with $n=3$ at the supersymmetric minimum. So the next procedure will be to check what happens to the structure of the potential density after adding (2.9) to the nonperturbative potential (2.8). Does the sum of these two potentials still respects (3.42) with $n=3$ at the dS minima? As an answer to this question, we will first show in Sec. 5.1 that $\delta {\mathcal V}_{\rm scalar}$, and consequently the sum of (2.8) and (2.9) does not respect (3.42). This means that the density of $\delta {\mathcal V}_{\rm scalar}$( $\equiv V_{\overline{D3}}$) caused by $\overline{D3}$-branes makes a nonzero contribution to $\lambda$ in (3.41), and consequently $\lambda$ of the dS vacua described by (2.8) plus (2.9) may not be fine-tuned to vanish unlike in the scenario of the original KKLT. Hence in the second part of this section (Sec. 5.2) we will propose an alternative scenario for the vanishing $\lambda$ of our present universe. This alternative scenario uses AdS, instead of dS, vacua of KKLT, and it has more nice properties as compared with those dS vacua uplifted by anti-$D3$-branes.

\vskip 0.3cm
\hspace{-0.6cm}{\bf \large 5.1 $\delta {\mathcal V}_{\rm scalar}$ due to $\overline{D3}$-branes}
\vskip 0.15cm

In \cite{15} the dynamics of $\overline{D3}$-branes is described by a Dirac-Born-Infeld (DBI) plus Chern-Simons (CS) world volume action for the $NS5$-brane due to technical difficulties in obtaining DBI action for the pure $\overline{D3}$-branes in the KS background geometry. In this S-dual description the $\overline{D3}$-branes are described by $NS5$-branes wrapping $S_2$ inside the $A$-cycle of the conifold geometry. At the apex of the conifold the metric becomes
\begin{equation}
ds^2 = a_0^2 dx_{\mu} dx^{\mu} + R_0^2 \big( d \psi^2 + \sin^2 \psi d \Omega_2^2 \big)\,\,,
\end{equation}
where $a_0$ and $R_0$ are constants, and the world volume action for the $NS5$-brane of type IIB theory takes the form (see \cite{15} or \cite{16})
\begin{equation}
I_{NS5} = \frac{\mu_5}{g_s^2} \int d^6 \xi \big[ -det (g_{\mu\nu}) \cdot det(h_{{\hat m}{\hat n}} + 2 \pi g_s {\mathcal F}_{(2)} ) \big]^{1/2} + \mu_5 \int B_{(6)} \,\,,
\end{equation}
where $h_{{\hat m}{\hat n}}$ is a two-dimensional metric induced along $S_2$ of the $A$-cycle and $2 \pi {\mathcal F}_{(2)} = 2\pi F_{(2)} -A_{(2)}$ with $F_{(2)} =dA$ a two-form field strength of the world volume gauge field of the $NS5$-brane. In (5.2) $F_{(2)}$ is assumed to satisfy
\begin{equation}
2 \pi \int_{S_2} F_{(2)} = 4 \pi^2 p \,\,,
\end{equation}
so that the $NS5$-brane carries $\overline{D3}$ charge $p$. R-R two-form $A_{(2)}$ is also assumed to satisfy
\begin{equation}
\int_{S_2} A_{(2)} = 4 \pi M \Big(\psi - \frac{1}{2} \sin (2\psi) \Big) \,\,,
\end{equation}
which follows from the well-known R-R flux quantization $\int_{A} F_{(3)} = 4 \pi^2 M$.

The DBI part of (5.2) contains an internal metric $h_{{\hat m}{\hat n}}$ because (5.2) is an world volume action for the $NS5$-brane rather than genuine $\overline{D3}$-brane. But using (5.3) and (5.4), one finds that (5.2) turns into
\begin{equation}
I_{\overline{D3}} = \int d^4 x \sqrt{- det (g_{\mu\nu})} \,\, \mathcal{L}_{\overline{D3}} (\psi)\,\,,
\end{equation}
which is typical of the world volume action for the $D3/\overline{D3}$-branes. In (5.5), $\mathcal{L}_{\overline{D3}} (\psi)$ can be written, upon taking ${\dot \psi}=0$, in the form
\begin{equation}
\mathcal{L}_{\overline{D3}} (\psi) = \frac{4 \pi^2 \mu_5 M}{g_s} {\hat V}(\psi)\,\,,
\end{equation}
where $M$ is related with $h_{{\hat m}{\hat n}}$ by the integral
\begin{equation}
\int_{S_2} d^2 y \sqrt{det (h_{{\hat m}{\hat n}} + 2 \pi g_s {\mathcal F}_{(2)})\,} \, = 4 \pi^2 M g_s  {\hat V}(\psi)\,\,,
\end{equation}
and ${\hat V}(\psi) \simeq p/M$ for $\psi \ll 1$ (see [16]). The scalar potential for the $\overline{D3}$-branes can be read from $\mathcal{L}_{\overline{D3}} (\psi)$ in (5.6) and it turns out to take the form (2.9).\footnote{To obtain (2.9) one should insert the scale factor $e^{2u(x)}$ of the internal space in the metric in advance.}

Turning back to (5.5), $\mathcal{L}_{\overline{D3}} (\psi)$ is given as a function of $\psi$ and where $\overline{D3}$-branes correspond to $\psi =0$. But in the KS geometry $\psi =0$ is not a stable, nor a metastable point of the potential, and hence in the S-dual description the $\overline{D3}$-branes are necessarily described by the $NS5$-branes which occupy $S_2$ of the $A$-cycle in the internal space. So $\mathcal{L}_{\overline{D3}} (\psi)$ in (5.5) necessarily contains the two-dimensional internal metric $h_{{\hat m}{\hat n}}$ implicitly in the form $\int d^2 y \sqrt{det (h_{{\hat m}{\hat n}} + 2 \pi g_s {\mathcal F}_{(2)})\,}$ (see (5.6) and (5.7)), and we infer that the potential density $V_{\overline{D3}}$ will be of the form
\begin{equation}
V_{\overline{D3}} \sim \frac{\mu_5}{g_s} \frac{\sqrt{det (h_{{\hat m}{\hat n}} + 2 \pi g_s {\mathcal F}_{(2)})\,}}{\sqrt{det (h_{{\hat m}{\hat n}})}} \,\delta^4 (y) \,\,,
\end{equation}
where $\delta^4 (y)$ is defined by $\int d^4 y \sqrt{h_4} \, \delta^4 (y) =1$ with $\sqrt{h_4} \equiv \sqrt{h_6}/ {\sqrt{det (h_{{\hat m}{\hat n}}) }}$. After all, we find that $V_{\overline{D3}} \notin V_n$ because $V_{\overline{D3}}$ in (5.8) does not satisfy (3.42).

\vskip 0.3cm
\hspace{-0.6cm}{\bf \large 5.2 dS vacua of KKLT and an alternative scenario}
\vskip 0.15cm

\vskip 0.3cm
\hspace{-0.6cm}{(1) dS vacua of KKLT}
\vskip 0.15cm

In KKLT, the scalar potential is so adjusted that the constituents (2.8) and (2.9) cancel out at the dS minima. So the scalar potentials of the dS vacua almost vanish at their dS minima $\sigma =\sigma_m$, and we can write
\begin{equation}
{\mathcal V}_{\rm dS} = \epsilon |{\mathcal V}_{\rm AdS}|\,\,,
\end{equation}
where $\epsilon$ is an arbitrarily small positive constant of order $\sim 10^{-120}/ O({\mathcal V}_{\rm AdS})$. If ${\mathcal V}_{\rm dS}$ in (5.9) can take sufficiently small values, the corresponding $\lambda$ will also be very small, and we may take one of the dS vacua of KKLT as the background vacuum of our present universe. However, this is true only in the traditional theories. According to our discussions in Sec. 5.1 it is very unlikely that such a fine-tuning is really possible.

The superpotentials of dS vacua do not satisfy $DW =0$ at the dS minimum because the introduction of anti-${D3}$-branes breaks the supersymmetry slightly. Thus the superpotentials for the dS vacua do not have the structure (4.9) or (4.17), and consequently ${\mathcal V}_{\rm dS}$ of dS vacua may not be able to be written in the form (4.20), which suggests that the corresponding $V_{\rm dS}$ necessarily makes a nonzero contribution to $\lambda$ in (3.41). Indeed in Sec. 5.1, we have shown that the density of $\delta {\mathcal V}_{\rm scalar}$ in (2.9) ($= V_{\overline{D3}}$) does not satisfy $V_{\overline{D3}} \in V_3$ and hence $\delta {\mathcal V}_{\rm scalar}$ arising from anti-$D3$-branes necessarily makes a nonzero contribution to $\lambda$ in the equation (3.41).

In the case of AdS vacua, however, it was shown that $V_{\rm AdS}$ belongs to $V_3$ (See Sec. IV.) and therefore it does not contribute to $\lambda$ as opposed to the case of $V_{\overline{D3}}$. So these things make us to doubt that $\lambda$ of dS vacua can be really fine-tuned to vanish by adding ${\mathcal V}_{\rm AdS}$ in (2.8) and $\delta {\mathcal V}_{\rm scalar}$ in (2.9). Indeed, even when we accept the possibility of this fine-tuning, it is preserved only at the tree level. The problem is that once the perturbations enter, there is no way to make $\lambda$ remain to be of order $\sim 10^{-120}$ in the units of Planck density. For instance if we take the quantum fluctuations on the $D3$-branes into account, the fine-tuning $\lambda =0$ will be severely disturbed. For these reasons it seems that we may need to introduce an alternative scenario which can substitute for the dS vacua of KKLT type models. In this section we propose a new vacuum scenario for the background state of our present universe, as a substitute for the original dS vacuum scenario of KKLT.

\vskip 0.3cm
\hspace{-0.6cm}{(2) AdS vacuum scenario}
\vskip 0.15cm

As mentioned above, in Sec. IV we have shown that the potential density of the AdS vacua belongs to $n=3$ ($V_{\rm AdS} \in V_3$) and consequently it does not make any nonzero contributions to $\lambda$. So the simplest, and perhaps the most natural scenario using KKLT is to take one of these AdS vacua of KKLT to identify it as the background vacuum of our present universe. (Recall that in our self-tuning mechanism ${\mathcal V}_{\rm scalar} <0$ does not necessarily imply $\lambda <0$ due to ${\mathcal E}_{\rm SB}$ in (1.1).) This AdS vacuum configuration with certain numbers of $D3$-branes may be identified with the supersymmetric (stable) minimum at $\psi =\pi$ of the brane/flux annihilation description in \cite{15}. Namely the nonsupersymmetric configuration with $p$ anti-${D3}$-branes (the dS vacua) rolls down (via tunneling and a classical process at some early stage during or after inflation) the potential to the north pole $\psi = \pi$ to form a supersymmetric configuration with $M-p$ $D3$-branes which is now identified with the present stage of our universe. In this scenario the supersymmetry breaking of the brane region is basically generated by ${\mathcal E}_{\rm SB}$, not by anti-${D3}$-branes (see Sec.VIII).

The above AdS vacuum scenario can substitute for the dS vacua of KKLT in the framework where $\lambda$ is given by (1.1), and AdS vacua of this scenario have more nice properties as compared with the dS vacua, as listed below.

\vskip 0.6cm
\begin{itemize}
\item In general dS vacua of the usual flux compactifications have a tunneling instability since these dS vacua are only local minima of the potential and they eventually decay into run away vacuum at $\sigma = \infty$. Hence in the theories using these dS vacua the authors need to show that their lifetimes are huge enough to describe our present universe as in KKLT. In the AdS vacuum scenario, however, the background (AdS) vacua describing our present universe are stable both classically and quantum mechanically and such a tunneling instability is inherently absent.
\item The dS vacua uplifted by anti-${D3}$-branes also suffer from another kind of tunneling instability. As mentioned above, nonsupersymmetric configurations with anti-${D3}$-branes (the dS vacua) correspond to the metastable states in the brane/flux annihilation descriptions in \cite{15}, and these metastable states decay, via tunneling and classical process, into supersymmetric configurations with ${D3}$-branes (the AdS vacua) which correspond to the stable minima of the brane/flux annihilation description. Since these AdS vacua correspond to the  stable minima, there is no other minimum (or minima) to decay into.
\item There is no any parameter, nor coefficient to be fine-tuned in the AdS vacuum scenario. $\lambda =0$ is automatically achieved by the cancelation between ${\mathcal V}_{\rm scalar} + \delta_Q {\hat I}_{\rm brane}^{(NS)} + \delta_Q {\hat I}_{\rm brane}^{(R)}$ and ${\mathcal E}_{\rm SB}$, forced by (3.41).
\item Most of all, in the AdS vacuum scenario of our self-tuning mechanism the fine-tuning $\lambda=0$ is radiatively stable. Any nonzero contributions to ${\mathcal V}_{\rm scalar}$ coming from $g_s$ - perturbations and quantum fluctuations (vacuum energies) on the visible sector $D3$-branes are all gauged away by ${\mathcal E}_{\rm SB}$ and as a result $\lambda =0$ is always preserved.
\end{itemize}

The dS vacua with anti-${D3}$-branes might be suitable for the description of the early universe including inflation, rather than the present universe with vanishing $\lambda$. The anti-${D3}$-branes are indispensable in the brane-antibrane inflation scenario \cite{17,18} because the potential for the inflation (inflaton potential) is generated by the brane-antibrane interaction. Also in the inflationary era the coefficient $D$ in (2.9) (and therefore $\epsilon$ in (5.9)) does not have to be fine-tuned. Entire ${\mathcal V}_{\rm dS}$ of the dS vacua can contribute, together with the potential generated by the brane-antibrane interaction, to $\lambda$ to make it positive. But these nonsupersymmetric dS vacua with anti-$D3$-branes are only metastable, hence they eventually decay into the supersymmetric AdS vacua describing our present universe in the AdS vacuum scenario.

\vskip 1cm
\hspace{-0.65cm}{\bf \Large VI. Open string moduli}
\vskip 0.5cm
\setcounter{equation}{0}
\renewcommand{\theequation}{6.\arabic{equation}}

In the AdS vacuum scenario the supersymmetric configuration at $\psi = \pi$ contains $D3$-branes in the KS throat. Introduction of $D3/\overline{D3}$-branes generally induces a scalar potential coming from the DBI plus CS action. For instance in KKLT, an introduction of anti-$D3$-branes induces an additional term (2.9) to the scalar potential as we have already seen. Also the potential for the $D3$-branes, which vanishes in the ISD compactifications, acquires nonzero contributions once the background turns into IASD because in this background the IASD fluxes become a source for the scalar potential of the $D3$-branes. Besides this, the presence of $D3/\overline{D3}$-branes also yields open string moduli such as locations of the branes in the compact space. Thus we may need to check if all these contributions to the scalar potential also respect (3.42) with $n=1,3$ to make $\lambda$ vanish. In this section we want to check the contributions coming from the open string moduli of the $D3$-branes, and then in the next section we will consider the $D3$-brane potential sourced by IASD fluxes. In our discussions of this section we will consider the general case where the nonperturbative vacua are basically given by the AdS type vacua, rather than dS, of KKLT according to the discussions of the previous section. So we do not have anti-${D3}$-branes in our configurations.

Suppose that we have a single (or a stack of) $D3$-brane(s) in the six-dimensional compact space for simplicity. In the presence of a $D3$-brane the K$\ddot{\rm a}$hler modulus\footnote{For simplicity we consider the configuration which only has a single K$\ddot{\rm a}$hler modulus as in KKLT.} $\rho$ acquires an additional term $k (Y, {\bar Y})$ \cite{19}:
\begin{equation}
\rho = \frac{b}{\sqrt 2} + i e^{4u} + \frac{i}{2} k (Y, {\bar Y}) \,\,,
\end{equation}
where the three complex scalars $Y^{\alpha}$, $\alpha =1,2,3$, in $k (Y, {\bar Y})$ represent the location of the $D3$-brane.\footnote{In our AdS vacuum scenario $D3$-branes are fixed at the apex of the Calabi-Yau cone, so in our case $Y^{\alpha}$ is simply $Y^{\alpha}=0$.} The K$\ddot{\rm a}$hler potential for this K$\ddot{\rm a}$hler modulus is therefore
\begin{equation}
{\mathcal K}_{\rho} = -3 \ln e^{4u} = -3 \ln \big[ -i ( \rho - {\bar \rho}) - k (Y, {\bar Y}) \big] \,\,.
\end{equation}
Besides this, the nonperturbative superpotential (2.1) also changes in the presence of $D3$-brane into the form \cite{19}
\begin{equation}
W=W_0 +A e^{i a \rho -\zeta (Y)} \,\,.
\end{equation}
So the supersymmetric vacua must satisfy
\begin{equation}
D_{\rho} W = i a A e^{i a \rho -\zeta (Y)} + \frac{3iW}{\big[ -i ( \rho - {\bar \rho}) - k (Y, {\bar Y}) \big]} =0 \,\,,
\end{equation}
\begin{equation}
D_{\alpha} W = - A \partial_\alpha \zeta (Y) e^{i a \rho -\zeta (Y)} + \frac{3 (\partial_\alpha k)W}{\big[ -i ( \rho - {\bar \rho}) - k (Y, {\bar Y}) \big]} =0 \,\,,
\end{equation}
and from these two equations one obtains
\begin{equation}
\partial_\alpha \zeta (Y) + a \partial_\alpha k(Y, {\bar Y}) =0 \,\,.
\end{equation}
(6.6) guarantees that (6.4) and (6.5) are not inconsistent with each other as far as it admits a solution.

Now we can show that the potential density $V_{\rm AdS}$ associated with the superpotential (6.3) still belongs to $V_3$. $W$ in (6.3) differs from $W$ in (2.1) only in that $e^{ia\rho}$ is replaced by $e^{ia\rho \, -\zeta(Y)}$, and in (6.3) the complex structure moduli are only contained in $W_0$ and $A$ as before. Also since the K$\ddot{\rm a}$hler potential for the complex structure moduli is still given by $\propto \ln [\,-i \int_{{\mathcal M}_6} \Omega \wedge {\bar \Omega}\,]$ (where $\Omega$ represents $\Omega(\tau^I )$), the F-term condition (4.8) requires $W$ to take the form (4.9) again except that $\bar \alpha$ may now depend on both $\tau$ and $Y^{\alpha}$, instead of $\tau$ alone. (But see the footnote 6.) Indeed, repeating the same procedure from eq. (4.10) to (4.16) one obtains (4.17) again. The only difference is that $\delta \tau^J$'s in $\tau_0^J + \delta \tau^J$ now also depend on $Y^{\alpha}$ in addition to $\tau_0^I$ and $\sigma_m$. So we finally obtain (4.20) again for ${\mathcal V}_{\rm AdS}$, implying that $V_{\rm AdS} \in V_3$ and therefore $V_{\rm AdS}$ does not contribute to $\lambda$ even in the presence of the open string moduli.

One can reaffirm the above result as follows. Substituting (6.3) into (6.4) gives
\begin{equation}
-3 \frac{W_0}{A} = \big[ 3- i a ( \rho - {\bar \rho}) - a k (Y, {\bar Y}) \big] e^{i a \rho -\zeta (Y)} \,\,.
\end{equation}
But since
\begin{equation}
- i ( \rho - {\bar \rho}) -  k (Y, {\bar Y}) = 2 e^{4u} \,\,
\end{equation}
from (6.1), one obtains
\begin{equation}
A = -W_0 e^{-i a \rho +\zeta (Y)} (1+\frac{2}{3} a e^{4u} )^{-1} \big|_m \,\,.
\end{equation}
(6.9) coincides with (4.22) except $e^{a \sigma_m}$ is replaced by $e^{-i a \rho +\zeta (Y)}|_{m}$. So ${\mathcal V}_{\rm AdS}$ obtained from (6.3), which will be identical with (2.8) only except that $e^{-a \sigma}$ is replaced by $e^{-a \sigma + \xi(Y)}$, will take the same form as (4.23) by (6.9) at the supersymmetric minimum, and by repeating the same discussions as in Sec. 4.2 one finds that $V_{\rm AdS}$ associated with (6.3) also belongs to $V_3$ as before.

\vskip 1cm
\hspace{-0.65cm}{\bf \Large VII. $D3$-brane potential}
\vskip 0.5cm
\setcounter{equation}{0}
\renewcommand{\theequation}{7.\arabic{equation}}

In the ISD compactifications $-$ and in the absence of branes $-$ $\lambda$ trivially vanishes from (3.41) because potential density arising from the fluxes vanishes in the ISD background. But once the perturbations come into the theory, ${\mathcal V}_{\rm scalar}$ does not vanish anymore because in this case $G_{(3)}$ aquires IASD components. Besides this, the IASD fluxes also induce a potential for the $D3$-branes because they become a dominant source in the equation of motion for the D3-brane potential. In \cite{20}, it was shown that there exist three distinct types of closed, IASD three-form fluxes which induce the $D3$-brane potential.

Among these fluxes the simplest one is the type I flux which contains only $G_{(1,2)}$, the IASD $G_{(3)}$ of Hodge type $(1,2)$. Compared with other two types of fluxes, the type I flux is of particular importance because the other two contain non-primitive $(2,1)$ which is forbidden in a compact Calabi-Yau space. Besides this, it was also shown in \cite{20} that there is a holographic correspondence between perturbations of supergravity solution by the type I flux and superpotential perturbations of the conformal field theory. In this correspondence the scalar potential for a probe D3-brane in the conifold geometry precisely matches the scalar potential computed in the gauge theory with superpotential $W$, and the scalar potential for a $D3$-brane in the conifold geometry is reproduced by the $G_{(1,2)}$ flux.

\vskip 0.3cm
\hspace{-0.6cm}{\bf \large 7.1 $D3/\overline{D3}$-brane potentials in the string frame}
\vskip 0.15cm

The $D3/\overline{D3}$-brane potentials follow from the DBI plus CS action (3.3) with $T(\phi) \backslash \mu(\phi)$ replaced by $T_3 e^{-\phi} \backslash \mu_3$. In string frame it is given by
\begin{equation}
I_{D3/\overline{D3}} = - T_3 \int d^4 x \, e^{-\phi} \sqrt{-det (g_{\mu\nu})} + \mu_3 \int A_{(4)} \,\,,
\end{equation}
where $T_3 = |\mu_3 | = (2\pi)^{-3} (\alpha^{\prime})^{-2}$ and
\begin{equation}
A_{(4)}= \xi(y) \sqrt{-g_4} \, dx^0 \wedge dx^1 \wedge dx^2 \wedge dx^3 \,\,.
\end{equation}
For the given compactification (3.4), $I_{D3/\overline{D3}}$ becomes
\begin{equation}
I_{D3/\overline{D3}} = - T_3 \int d^4 x \sqrt{-g_4} \, \frac{1}{g_s} \Phi_{\mp}\,\,,
\end{equation}
where $\Phi_{\pm}$ are defined by
\begin{equation}
 \frac{1}{g_s} \Phi_{\pm} = \frac{\chi^{1/2}}{g_s} \pm \xi \,\,.
\end{equation}
Here we ignored the kinetic terms of the $D3/\overline{D3}$-brane actions because we assumed that the $D3/\overline{D3}$-branes are all fixed at some certain points of the compact space. According to (7.4), $\Phi_{-}$ vanishes in the ISD background if the (bulk) supersymmetry is unbroken (see Sec. V of \cite{1}), and therefore $D3$-branes feel no potential in this case. But once the higher order perturbations come into the theory, the situation changes. Because higher order terms of $G_{(3)}$ generally contain IASD components, and these components become a dominant source in the equation of motion for $\Phi_{-}(y)$ \cite{19, 20}, the $D3$-branes certainly feel a potential arising from the higher order terms of $\Phi_{-}(y)$. In this section we will show that this $D3$-brane potential arising from the IASD flux perturbations also respects the condition (3.42), but this time not with $n=3$, but with $n=1$. Namely $V_{D3} \in V_1$ for the type I flux.

\vskip 0.3cm
\hspace{-0.6cm}{\bf \large 7.2 Equation of motion for $\Phi_{-}$}
\vskip 0.15cm

The equation of motion for $\Phi_{-}$ may be obtained from the field equations for $\chi^{1/2}$ and $\xi$, among which the latter follows from the field equation for $A_{(4)}$. The field equation for $A_{(4)}$ can be obtained from the three terms
\begin{equation}
\frac{1}{8 \kappa_{10}^2} \int {\tilde F}_{(5)} \wedge \ast {\tilde F}_{(5)} + \frac{1}{8i \kappa_{10}^2} \int e^{\phi} A_{(4)} \wedge G_{(3)} \wedge {\bar G}_{(3)} + \frac{\mu_3}{2} \int A_{(4)}
\end{equation}
in the actions (3.1) and (3.3), where we have rewritten the ${\tilde F}_{(5)}^2$ term in (3.1) as $\frac{1}{8 \kappa_{10}^2} \int {\tilde F}_{(5)} \wedge \ast {\tilde F}_{(5)}$ for convenience, and replaced $\mu_3 \rightarrow \frac{\mu_3}{2}$ which is necessary to obtain correct equation for the self-dual field $A_{(4)}$ (see for instance \cite{14} or \cite{20-1} for this). We obtain from (7.5)
\begin{equation}
d \ast {\tilde F}_{(5)} = \frac{G_3 \wedge {\bar G}_3}{2 i Im \tau} + 2 \kappa_{10}^2 \,\mu_3 \,\rho_3^{\rm loc} \,\,,
\end{equation}
which, by (3.2), reduces to
\begin{equation}
\nabla^2 \xi = \frac{i}{12 Im \tau}\, \chi \, G_{mnp} \ast_6 {\bar G}^{mnp} + 2 \chi^{-1/2} (\partial \chi^{1/2}) (\partial \xi) + 2 \kappa_{10}^2 \mu_3 \, \chi \, \rho_3^{\rm loc} \,\,.
\end{equation}

The field equation for $\chi^{1/2}$, on the other hand, can be obtained from (3.6) plus the topological term
\begin{equation}
\frac{1}{8i \kappa_{10}^2} \int e^{\phi} A_{(4)} \wedge G_{(3)} \wedge {\bar G}_{(3)}=\frac{1}{2 \kappa_{10}^2 g_s^2} \big[\int d^4 x \sqrt{-g_4}\,\big] \Big(\frac{i g_s^2}{24} \int d^6 y \sqrt{h_6} \,e^{\phi} \xi \,G_{mnp} \ast_6 {\bar G}^{mnp}\Big)\,\,.
\end{equation}
Varying the action with respect to $B$ we obtain
\begin{equation}
\nabla^2 \Big( \frac{\chi^{1/2}}{g_s} \Big) = \frac{i}{12 Im \tau}\, \chi \, G_{mnp} \ast_6 {\bar G}^{mnp} +
\frac{1}{6 Im \tau} \, \chi \, G_{mnp}^{+}{\bar G}^{+ mnp} + \Big( \frac{\chi^{1/2}}{g_s} \Big)^{-1} \Big[ \partial \Big( \frac{\chi^{1/2}}{g_s} \Big) \Big]^2 \nonumber
\end{equation}
\begin{equation}
+ \Big( \frac{\chi^{1/2}}{g_s} \Big)^{-1} (\partial \xi)^2 + \frac{\beta}{g_s} + 2 \kappa_{10}^2 T_3 \, \chi \, \rho_3^{\rm loc} \,\,.
\end{equation}
Finally the equation of motion for $\Phi_{-}$ can be obtained by subtracting (7.7) from (7.9). Upon setting $\mu_3 =T_3$, we obtain
\begin{equation}
 \nabla^2 \Phi_{-} = \frac{g_s}{6 Im \tau} \, \chi \, |G_{(3)}^{+}|^2 + {\chi^{-1/2}} |\partial \Phi_{-}|^2 +  \beta \,\,.
\end{equation}
(7.10) is the string frame version of Eq. (2.8) of \cite{20}, and they coincide if we replace $\chi^{1/2}$ by $e^{4A}$, and $h^{mn}$ by $e^{-\phi /2}h^{mn}$.

\vskip 0.3cm
\hspace{-0.6cm}{\bf \large 7.3 IASD three-form fluxes}
\vskip 0.15cm

(7.10) shows that the IASD fluxes $G_{(3)}^{+}$ become a source for the potential $\Phi_{-}$. The explicit forms of the IASD fluxes can be found systematically by solving the equation of motion \cite{6,20}
\begin{equation}
d\Lambda + \frac{i}{Im \tau} d\tau \wedge Re \Lambda =0 \,\,
\end{equation}
perturbatively around ISD solutions. (7.11) can be obtained from a linear combination of the field equations for $A_{(2)}$ and $B_{(2)}$, and where $\Lambda$ is defined by
\begin{equation}
\Lambda = \Phi_{+} G_{-} + \Phi_{-} G_{+}\,\,, ~~~~~~ \big( G_{\pm} \equiv \pm i G_{(3)}^{\mp}  \big) \,\,.
\end{equation}
To solve (7.10) perturbatively we expand all fields as \cite{20}
\begin{equation}
X = X_{(0)} + X_{(1)} + X_{(2)} + \cdots \,\,,
\end{equation}
where $X_{(0)}$ represents the background fields and in particular $\Phi_{-}$ and $G_{-}$ both vanish in the ISD background
\begin{equation}
\Phi_{-}^{(0)} = G_{-}^{(0)}=0 \,\,.
\end{equation}
Since $\Lambda_{(0)}=0$ by (7.14), we need to solve (7.11) for $\Lambda_{(1)}$ which is now given by
\begin{equation}
\Lambda_{(1)} = \Phi_{+}^{(0)} G_{-}^{(1)} + \Phi_{-}^{(1)} G_{+}^{(0)}
\end{equation}
from (7.12) and (7.14).

At first order, (7.11) reduces to \cite{20}
\begin{equation}
d \Lambda_{(1)} = 0\,\,,
\end{equation}
which requires that $\Lambda_{(1)}$ should be a closed three-form. Also in (7.10) the flux-induced $\Phi_{-}$ should be of the second order because the smallest order of nonvanishing $G_{-}$ is already first order by (7.14). Hence we put $\Phi_{-}^{(1)} =0$ and therefore $\Lambda_{(1)}=\Phi_{+}^{(0)} G_{-}^{(1)}$ from (7.15), which shows that $\Lambda_{(1)}$ is IASD \cite{20},
\begin{equation}
\ast_{6}^{(0)} \Lambda_{(1)} = -i \Lambda_{(1)} \,\,,
\end{equation}
in the background metric. Finally for $\Phi_{-}^{(1)} =0$, (7.10) reduces to\footnote{In (7.18) we have set $\beta =0$ because we are considering the present stage (with $\lambda=0$) of our universe.} \cite{20}
\begin{equation}
 \nabla^2 \Phi_{-} = \frac{g_s^2}{24} \big| \Lambda \big|^2 \,\,,
\end{equation}
where $\Phi_{-}$, $\Lambda$ and $ \nabla^2$ are $\Phi_{-} = \Phi_{-}^{(2)}$, $\Lambda = \Lambda_{(1)}$ and $\nabla^2 =  \nabla_{(0)}^2$, respectively. So the potential $\Phi_{-}$ arising from the IASD fluxes can be obtained from (7.18) if we know the explicit forms of $\Lambda$ which is any closed, IASD three-form allowed on the Calabi-Yau cones.

Fortunately, the explicit solutions for the flux perturbations on arbitrary Calabi-Yau cones have been thoroughly studied in \cite{20}. According to the computations of \cite{20} there exist three distinct types of closed, IASD three-forms. See Sec. 3.3.2 of \cite{20} for these three types of IASD three-forms. Among these fluxes the type I flux is of particular importance since its contribution to $\Phi_{-}$ is dominant over the other two in the neighborhood of $y=0$ where the visible sector $D3$-branes are located. (We will see this soon.) Also the type II and III fluxes contain non-primitive $(2,1)$ which is forbidden in a compact Calabi-Yau space.\footnote{But for chiral perturbations, each flux becomes of pure Hodge type, and the type II and III fluxes do not contain non-primitive $(2,1)$ anymore.}

The potential $\Phi_{-}$ due to the type I flux is found to be \cite{20}
\begin{equation}
\Phi_{-} = \frac{g_s^2}{8} h^{\alpha {\bar \alpha}} \nabla_{\alpha}{f}_1 {\overline{\nabla_{\alpha}{f}_1}} \,\,,
\end{equation}
which is an $F$-term potential due to the superpotential perturbations of the form $\int d^2 \theta \, \triangle W$ with $\triangle W \sim f_1$. (7.19) suggests that the potential density $V_{D3}$ induced by the type I flux belongs to $V_1$. Indeed from (7.3) and (3.17) $V_{D3}$ can be written as
\begin{equation}
V_{D3} = 2 \kappa_{10}^2 g_s T_3 \Phi_{-}(0) \delta^6 (y) \,\,,
\end{equation}
and since $\Phi_{-}$ in (7.19) contains a single $h^{mn}$ in the real basis, (7.20) shows that $V_{D3} \in V_1$. So $V_{D3}$ induced by the type I flux does not contribute to ${\mathcal V}_{\rm scalar}$ in $\lambda$ (see (3.43)).

\vskip 0.3cm
\hspace{-0.6cm}{\bf \large 7.4 $D3$-branes located at $y=0$}
\vskip 0.15cm

In (7.20) $V_{D3}$ is proportional to $\Phi_{-} (0)$ instead of $\Phi_{-} (y)$, which is due to the fact that in our AdS vacuum scenario the $D3$-branes are not the mobile branes anymore because we are considering the present (not inflationary) stage of our universe. In our descriptions of the present universe (a stack of) visible sector $D3$-branes are assumed to be fixed at the apex $y=0$ of the Calabi-Yau cones, and consequently we have the delta-function $\delta^6 (y)$ in (7.20), and also $\Phi_{-} (0)$ instead of $\Phi_{-} (y)$. The presence of delta-function, or having $\Phi_{-} (0)$ instead of $\Phi_{-} (y)$ in $V_{D3}$ enables us to ignore the whole (not just only the type I) contributions to $V_{D3}$ arising from the above three types of flux perturbations. This can be shown as follows.

The field equation (7.18) can be solved by using the green function method. Again in \cite{20} it was found that the resulting spectrum of $\Phi_{-}$ can be written as\footnote{We would like to thank the authors of \cite{20} for presenting very useful results of the complete studies on the issue under discussion.}
\begin{equation}
\Phi_{-} (y) = \sum_{\delta_{i}, \delta_{j}} r^{\triangle ({\delta_{i}, \delta_{j}})} h_{({\delta_{i}, \delta_{j}})} (\Psi) \,\,,
\end{equation}
where $h_{({\delta_{i}, \delta_{j}})} (\Psi)$ are angular wave functions which are related to the harmonics $Y_{LM} (\Psi)$ of the unperturbed Laplacian and $\triangle ({\delta_{i}, \delta_{j}})$ are radial scaling dimensions defined by
\begin{equation}
\triangle = \delta_{i}+ \delta_{j} -4 \,\,,
\end{equation}
where $\delta_{i}$ and $\delta_{j}$ are the scaling dimensions of the fluxes $\Lambda_i$ and $\Lambda_j$. The smallest value of $\triangle$ is obtained from a square of $\delta = \frac{5}{2}$ chiral mode of the type I flux, for which $\Phi_{-}$ is linear in $r$, $\Phi_{-} \propto r$. The other smallest scaling dimensions (including the above $\triangle =1$ of $\delta_1 = \delta_2 = \frac{5}{2}$) of the flux-induced potential are (See Sec. 4.1.3 of \cite{20}.)
\begin{equation}
\triangle = 1, \, 2, \, \frac{5}{2}, \, \sqrt{28}-\frac{5}{2}, \, \cdots,
\end{equation}
which shows that the contribution of the type I flux to $\Phi_{-}$ is dominant over the other two as $r \rightarrow 0$. In any case, every term in (7.21) vanishes at $r=0$ for any $\triangle ({\delta_{i}, \delta_{j}})$, and so does $\Phi_{-} (0)$ in (7.20) as well. This suggests that the contributions of the other two can be also ignored $-$ despite that they do not belong to $V_n$ with $n=1,3$ $-$ since $\Phi_{-} (0)$, and therefore $V_{D3}$ itself vanishes for the $D3$-branes fixed at $r=0$ of the Calabi-Yau cones.

Apart from this, the potential $\Phi_{-}$ can also include harmonic functions on the cones as the homogeneous solutions to (7.18). The contributions of these harmonic functions, however, can be also ignored for the $D3$-branes fixed at $r=0$. The harmonic expansion performed on the conifold takes the form \cite{21}
\begin{equation}
f(r, \Psi) = \sum_{L,M} c_{_{LM}} \Big( \frac{r}{r_{_{UV}}} \Big)^{\triangle_f (L)} Y_{LM} (\Psi) +c.c. \,\,, ~~~~~(r < r_{_{UV}})\,\,,
\end{equation}
where $c_{_{LM}}$ are constant coefficients and the radial scaling dimensions $\triangle_f (L)$ take the values of
\begin{equation}
\triangle_f (L) = \frac{3}{2}, \, 2, \, 2, \, 3, \, \sqrt{28}-2, \, \cdots \,\,.
\end{equation}
Since $\triangle_f (L)$ are all positive, all terms in (7.24) vanish at $r=0$, and therefore we can also ignore these contributions of the harmonic functions to $V_{\rm D3}$ as well.

In addition to these terms there might be a constant term, which is the trivial solution to the Laplace equation $\nabla^2 f=0$. This constant term does not vanish at $r=0$. However, it might be irrelevant to our configurations which do not involve $\overline{D3}$-branes. The constant term appears in the perturbative expansion of the $\overline{D3}$-brane potential $T_3 \Phi_{+} (r;r_{0})$ ($\equiv V_{D3/\overline{D3}} (r)$) (see \cite{18}). Since mobile $D3$-branes affect $\Phi_{+}$ perturbatively, $V_{D3/\overline{D3}}$ depends on the $D3$-brane position $r$, and it serves as a potential for the $D3$-brane. In this expansion of $V_{D3/\overline{D3}} (r)$ the constant term appears as the unperturbed potential energy of the $\overline{D3}$-branes fixed at $r=r_0$, and therefore it must vanish for the configurations which do not contain $\overline{D3}$-branes. After all, those terms (including harmonic functions) arising from the Coulomb interaction $V_{D3/\overline{D3}}$ between $D3$-branes and $\overline{D3}$-branes must all be excluded from $V_{D3}$ since they are irrelevant to our AdS vacuum scenario which does not involve the $\overline{D3}$-branes at all. See Sec. 5.2.

Besides all this, we finally observe that (3.41) contains the factor $\chi^{1/2}$. In the simple compactifications with $F_{(3)}=H_{(3)}=0$, $\chi^{1/2}$ takes the form (see eq. (5.8) of \cite{1})
\begin{equation}
\chi^{1/2} (r) = \Big( 1+ \frac{Q_0}{r^4} \Big)^{-1} \,\,,~~~~~\Big( Q_0 \equiv \frac{2 \kappa_{10}^2 g_s \mu_3}{4 \rm Vol(B)}\, \Big)\,\,,
\end{equation}
and in the neighborhood of $y=0$ it becomes $\chi^{1/2} (r) \sim {r^4} / {Q_0}$.\footnote{$Q_0$ in $\chi^{1/2}$ will cancel with $2 \kappa_{10}^2 g_s T_3$ in $V_{\rm D3}$ (see (7.20)) in the self-tuning equation (3.41).} Hence the densities $V$ which survive the projection $\Pi_\lambda ({\mathcal N}) \equiv  \frac{1}{24} \, \chi^{1/2} ({\mathcal N} - 1)({\mathcal N} -3) (1-3 b_0 \Pi ({\mathcal N}))$ in (3.41) (i.e. those $V$'s that do not respect (3.42) with $n=1,3$ just like $V_{D3}$ due to the type II and III IASD fluxes for instance) are highly suppressed again because they all have an extra factor $\chi^{1/2} (r)$ which strongly vanishes at $y=0$ ($r=0$) in the approximation $F_{(3)}=H_{(3)}=0$.\footnote{The compactifications with $F_{(3)}=H_{(3)}=0$ are good approximations in the AdS vacuum scenario because in KKLT the superpotential $W_0$ (and therefore $G_{(3)}$) is only of an order $\sim 10^{-4}$, instead of $\sim O(1)$.}

\vskip 1cm
\hspace{-0.65cm}{\bf \Large VIII. Summary and Discussion}
\vskip 0.5cm
\setcounter{equation}{0}
\renewcommand{\theequation}{8.\arabic{equation}}

\vskip 0.3cm
\hspace{-0.6cm}{\bf \large 8.1 Summary of our self-tuning mechanism}
\vskip 0.15cm

In an attempt to address the cosmological constant problem (especially aiming at explaining the fine-tuning $\lambda=0$ of our present universe) we have considered a new type of self-tuning mechanism whose basic principle has been partially presented in \cite{1}. The main point of this self-tuning mechanism can be summarized as
\begin{itemize}
\item Whether $\lambda$ vanishes or not is basically determined (in the six-dimensional internal space) by the tensor structure of the scalar potential density $V$, not by the zero or nonzero values of the scalar potential ${\mathcal V}_{\rm scalar}$ itself. If the density of ${\mathcal V}_{\rm scalar}$ belongs to one of $V_n$ with $n=1,3$, then $\lambda$ is forced to be fine-tuned to vanish regardless of whether ${\mathcal V}_{\rm scalar}$ vanishes or not.
\item In the new self-tuning mechanism $\lambda$ contains an exceptional term ${\mathcal E}_{\rm SB}$, and this ${\mathcal E}_{\rm SB}$ has its own gauge arbitrariness. So any nonzero ${\mathcal V}_{\rm scalar}$ and quantum fluctuations $\delta_Q {\hat I}_{\rm brane}^{(NS)} + \delta_Q {\hat I}_{\rm brane}^{(R)}$  on the branes can be gauged away by this ${\mathcal E}_{\rm SB}$ so that $\lambda$ vanishes as a result. The cancelation between ${\mathcal V}_{\rm scalar} +\delta_Q {\hat I}_{\rm brane}^{(NS)} + \delta_Q {\hat I}_{\rm brane}^{(R)}$ and ${\mathcal E}_{\rm SB}$ is automatically achieved by a self-tuning equation (3.41) once the density of ${\mathcal V}_{\rm scalar}$ satisfies $V \in V_n$ with $n=1,3$ as stated above.
\item Hence in the new self-tuning mechanism the self-tuning $\lambda=0$ is radiatively stable. Any contributions to ${\mathcal V}_{\rm scalar}$ coming from $g_s$-perturbation and quantum fluctuations on the $D3$-branes are all gauged away by ${\mathcal E}_{\rm SB}$ (and by a self-tuning equation), and as a result $\lambda=0$ is always preserved as mentioned above.
\end{itemize}

We applied the above self-tuning mechanism to the well-known scenario of KKLT to obtain a realistic model of our present universe with nearly vanishing cosmological constant. As a result of this application we found that the simplest, and perhaps the most natural scenario using KKLT is to take one of the AdS, instead of dS, vacua of KKLT as the background vacuum of our present universe. These AdS vacua are stable both classically and quantum mechanically. They do not have the tunneling instabilities of the dS vacua uplifted by anti-${D3}$-branes. The AdS vacuum scenario suggests that the F-term upliftings in the literature \cite{2,22} are basically unnecessary in obtaining a vanishing (or a nearly-vanishing) cosmological constant. The vanishing $\lambda$ is automatically achieved by the self-tuning equation (3.41), and by the gauge arbitrariness of ${\mathcal E}_{\rm SB}$ contained in (1.1). Namely the AdS vacuum scenario, or the self-tuning mechanism of this paper is basically realized by the two unusual equations (1.1) and (3.41).

The first equation (1.1) suggests that the cosmological constant $\lambda$ is not simply given by a scalar potential ${\mathcal V}_{\rm scalar}$ alone. According to (1.1), $\lambda$ contains an additional term, the supersymmetry breaking term ${\mathcal E}_{\rm SB}$, which possesses its own gauge arbitrariness. Hence in our case $\lambda=0$ does not necessarily imply ${\mathcal V}_{\rm scalar}=0$, and AdS vacua with ${\mathcal V}_{\rm scalar}<0$ are not inconsistent with $\lambda =0$ unlike in the theories where $\lambda$ is directly identified with ${\mathcal V}_{\rm scalar}$. In our self-tuning mechanism $\lambda$ is generally given by (3.20). But in the AdS vacuum scenario proposed in Sec. 5.2, $V$ is basically given by $V_{\rm AdS}$, and in this case (3.20) reduces (upon using (3.17)) back to (1.1) by (3.40) and (3.45) because $V_{\rm AdS} \in V_3$ and therefore $({\mathcal N}-1) V_{\rm AdS} = 2V_{\rm AdS}$. This result does not change even when we add $V_1$ (for instance, $V_{\rm D3}$ due to $\Phi_{-}$ in (7.19)) to $V$ because $({\mathcal N}-1) V _1$ simply vanishes.

Together with (1.1), the second equation (3.41) suggests that whether $\lambda$ vanishes or not is basically determined by the tensor structure of the potential density $V$, not by the zero or nonzero values of ${\mathcal V}_{\rm scalar}$ itself. (3.41) leads to the self-tuning $\lambda=0$ once our $V$ belongs to one of the class $V_n$ with $n=1,3$. We have shown that the AdS vacua of KKLT (including open string moduli of $D3$-branes) belong to $V_3$. So $\lambda$ of our present universe must tune itself to zero in the AdS vacuum scenario of Sec. 5.2. The negative values of ${\mathcal V}_{\rm scalar}$ of the AdS vacua are gauged away by ${\mathcal E}_{\rm SB}$ in (1.1), and $\lambda =0$ is automatically achieved by (3.41). This self-tuning process is not affected by the perturbations ${\mathcal K}= {\mathcal K}_{\rm tree} + {\mathcal K}_{\rm p} + {\mathcal K}_{\rm np}$ and $W= W_{\rm tree} + W_{\rm np}$ because these perturbations do not change the tensor structure of $V$. (See the last paragraph of Sec. IV.) Thus the whole radiative corrections of ${\mathcal V}_{\rm scalar}$ are also gauged away by (3.41), and the fine-tuning $\lambda =0$ remains stable against these corrections in the self-tuning mechanism of this paper.

The background vacua of the AdS vacuum scenario are supersymmetric, and therefore stable unlike the dS vacua uplifted by anti-${D3}$-branes. In the descriptions in \cite{15} the dS vacua necessarily involve the anti-$D3$-branes. So they are not supersymmetric and they are stable only classically at most. The dS vacua must eventually decay into the supersymmetric configurations of AdS vacua by the brane/flux annihilation process of \cite{15}, and this also suggests that the AdS vacuum scenario is more natural description of our present universe as compared with the dS vacua uplifted by anti-${D3}$-branes.

\vskip 0.3cm
\hspace{-0.6cm}{\bf \large 8.2 Supersymmetry breaking in the AdS vacuum scenario}
\vskip 0.15cm

In the AdS vacuum scenario the supersymmetry is basically broken by ${\mathcal E}_{\rm SB}$ in (1.1) (and by IASD components of the three-form fluxes arising from the perturbations), not by anti-$D3$-branes. In order to see it we rewrite (3.3) plus (3.39) as
\begin{equation}
I_{\rm brane}= \Big[ \int d^4 x \sqrt{-g_4} \, \Big] \int r^5 dr \epsilon_5 \Big( - e^{2B} T(\phi) + \mu (\phi) \xi (r) + \delta \mu_{T}^m (\phi) f_m (y) \Big) \delta^6 (r) \,\,,
\end{equation}
where $T(\phi)$ and $\mu (\phi)$ are given by
\begin{equation}
T(\phi) = T_3 e^{-\phi} + \rho_{\rm vac} (\phi)\,\,,~~~~~\mu (\phi) = \mu_3 + \delta \mu (\phi)\,\,,
\end{equation}
because we are now taking quantum fluctuations on the branes into account. The last term of (8.1) occurs as a result of the gauge symmetry breaking of $A_{(4)}$ arising at the quantum level. The substance of this term is a vacuum energy density of the brane region arising from the quantum excitations with components along the transverse directions to the $D3$-branes, and it plays very important roles in the supersymmetry breaking of the brane region, and in the process of self-tuning $\lambda =0$.

\vskip 0.3cm
\hspace{-0.6cm}{(1) With vanishing three-form fluxes}
\vskip 0.15cm

The supersymmetry transformations of the fermi fields of type IIB supergravity are \cite{23}
\begin{equation}
\delta \chi_{\phi} = \frac{1}{2} \Gamma^m (\partial_m \phi) \eta + \frac{i}{4} e^{\phi}\, {\bar {\mathbf G}}^{(3)} \eta^{\ast} \,\,,
\end{equation}
\begin{equation}
\delta \psi_m = \nabla_m \eta + \frac{i}{16} e^{\phi} \, {\tilde {\mathbf F}}^{(5)} \Gamma_m \eta - \frac{1}{8} \big(2 {\mathbf H}_m^{(3)} +i e^{\phi} \,{\mathbf F}^{(3)} \Gamma_m \big) \eta^{\ast} \,\,,
\end{equation}
where ${\mathbf F}^{(n)}$, ${\mathbf F}_m^{(n)}$ are defined by
\begin{equation}
{\mathbf F}^{(n)} \equiv \frac{1}{n!} \,\Gamma^{M_{1} \cdots M_{n}} F_{M_{1} \cdots M_{n}}\,\,,~~~~~{\mathbf F}_m^{(n)}\equiv \frac{1}{(n-1)!} \, \Gamma^{M_{1} \cdots M_{n-1}} F_{m M_{1} \cdots M_{n-1}}\,\,.
\end{equation}
In (8.3) and (8.4), the last terms represent supersymmetry transformations generated by the three-form fluxes $F_{(3)}$ and $H_{(3)}$. But if we want an easy understanding of the supersymmetry breaking of the AdS background, it is useful to consider a simple situation where the three-form fluxes are turned off, $F_{(3)}=H_{(3)}=0$ or $G_{(3)}=0$. Note that such compactification is a good approximation in our AdS vacuum scenario because in KKLT the superpotential $W_0$ (and therefore $G_{(3)}$) is only of an order $\sim 10^{-4}$ (see footnote 11). So the supersymmetry breaking  generated by the three-form fluxes $-$ regardless of whether ISD or IASD $-$ could be neglected for a moment in the simplified analysis for the core principle. For $F_{(3)}=H_{(3)}=0$, (8.3) and (8.4) reduce to
\begin{equation}
\delta \chi_{\phi} = \frac{1}{2} \Gamma^m (\partial_m \phi) \eta \,\,, ~~~~~~\delta \psi_m = \nabla_m \eta + \frac{i}{16} e^{\phi} \, {\tilde {\mathbf F}}^{(5)} \Gamma_m \eta\,\,,
\end{equation}
and these $\delta \chi_{\phi}$ and $\delta \psi_m$ vanish for (7.26) and constant $\phi$ \cite{24}. Hence in the approximation $F_{(3)}=H_{(3)}=0$, the supersymmetry is unbroken when $\phi$ is constant \cite{1}.

Now consider the field equation for $\phi$. Using (8.1) we obtain
\begin{equation}
\nabla^2 \phi - \frac{i g_s^2}{12} \Big( \frac{\Phi_{-}}{g_s} \Big) e^{\phi} G_{mnp} \ast_6 {\bar G}^{mnp} - \frac{g_s^2}{6} \Big( \frac{\chi^{1/2}}{g_s} \Big) e^{\phi} G^{+}_{mnp}{\bar G}^{+mnp} ~~~~~~~~~~~~~~~~~~~~\nonumber
\end{equation}
\begin{equation}
~~~~~= 2 \kappa_{10}^2 g_s^2 \Bigg[ e^{2B} \Big( T(\phi) + \frac{\partial T(\phi)}{\partial \phi} \Big) -  \frac{\partial \mu(\phi)}{\partial \phi} \xi (r)- \frac{\partial \delta \mu_{T}^{m}(\phi)}{\partial \phi} f_m (y) \Bigg] \delta^6 ({\vec r})
\end{equation}
from a linear combination of the field equations for $\hat{\phi}$ and $B$. In the given approximation (8.7) reduces to
\begin{equation}
\nabla^2 \phi =0
\end{equation}
in the bulk region, and therefore the bulk supersymmetry remains unbroken because (8.8) admits constant solutions. In the brane region, however, (8.7) reduces to
\begin{equation}
\nabla^2 \phi = c_0 \chi^{1/2} e^{\phi} \Big( \rho_{\rm vac} + \frac{\partial \rho_{\rm vac}}{\partial \phi} \Big) - c_0 g_s \Big(  \frac{\partial \delta \mu} {\partial \phi} \xi + \frac{\partial \delta \mu_{T}^{m}} {\partial \phi} f_m \Big) \,\,,
\end{equation}
where $c_0 = 2 \kappa_{10}^2 g_s \delta_0$ (see eq. (6.12) of \cite{1}), and $\rho_{\rm vac}$, $\delta \mu$, $\delta \mu_T^m$ are expanded respectively as
\begin{equation}
\rho_{\rm vac} (\phi) = \sum_{n=0}^{\infty} \rho_{(n)}\, e^{n \phi} \,\,,~~~~~\delta \mu (\phi) = \sum_{n=1}^{\infty} \mu_{(n)}\, e^{n \phi} \,\,, ~~~~~\delta \mu_T^m (\phi) = \sum_{n=1}^{\infty} \nu_{(n)}^m  \, e^{n \phi} \,\,.
\end{equation}
In \cite{1} it was shown (up to one-loop level) that all but the last term in (8.9) cancel out for $\mu_3 =T_3$ and $\mu_{(1)} = \rho_{(0)}$ which are required by consistency equations (see Sec. VIC of \cite{1}), and we are left with
\begin{equation}
\nabla^2 \phi =- c_0 \rho_{T}^{(1)} \,\,,~~~~~~ \big( \rho_{T}^{(1)} = \nu_{(1)}^m f_m \big) \,\,,
\end{equation}
and similarly we obtain
\begin{equation}
{\hat I}_{\rm brane} = \delta_0 \int r^5 dr \epsilon_5 \rho_{T}^{(1)}
\end{equation}
from (8.1) (see eq. (7.2) of \cite{1}). Since (8.12) comes from the last term of (8.1), it is identified (at one-loop level) with $\delta_G {\hat I}_{\rm brane}^{(R)}$, or equivalently with $-{\mathcal E}_{\rm SB}$ by (3.45) (see (1.2)). In (8.11), $\rho_{T}^{(1)}$ sources the supersymmetry breaking of the brane region, and ${\mathcal E}_{\rm SB}$ in (1.2) plays the role of a supersymmetry breaking term because the term $-c_0 \rho_{T}^{(1)}$ in (8.11) is obtained from the last term $\delta_G {\hat I}_{\rm brane}^{(R)} (=-{\mathcal E}_{\rm SB})$ of the action ${\hat I}_{\rm brane}$. Hence in the brane region the supersymmetry is broken by $\rho_{T}^{(1)}$ even in the absence of three-form fluxes, while in the bulk region it remains unbroken in that approximation.

Turning back to (1.1), ${\mathcal V}_{\rm scalar}$ of the AdS vacua vanishes for $G_{(3)}=0$ because the supersymmetric AdS vacua are defined by $DW =0$, which then implies ${\mathcal V}_{\rm AdS} \propto |W_{0}|^2$ from (2.4) and (4.22) (see (4.23)), and therefore ${\mathcal V}_{\rm AdS}$ vanishes for $G_{(3)}=0$ because so does $W_0$. Hence in the absence of the three-form fluxes (1.1) becomes
\begin{equation}
\lambda = \frac{\kappa^2}{2} \big( \delta_Q {\hat I}_{\rm brane}^{(NS)} + \delta_Q {\hat I}_{\rm brane}^{(R)} -{\mathcal E}_{\rm SB} \big)\,\,,
\end{equation}
and from (8.13) the self-tuning $\lambda=0$ requires that the energy scale ${\mathcal E}_{\rm SB}$ of the supersymmetry breaking must be equal to the magnitude of the non-vanishing fluctuations $\delta_Q {\hat I}_{\rm brane}^{(NS)} + \delta_Q {\hat I}_{\rm brane}^{(R)}$ on the branes. Also if the branes are BPS $D3$-branes with unbroken supersymmetry (which is the simplest, but not realistic case), $\delta_Q {\hat I}_{\rm brane}^{(NS)} + \delta_Q {\hat I}_{\rm brane}^{(R)}$ is expected to cancel out and $\lambda$ is simply given by
\begin{equation}
\lambda = \frac{\kappa^2}{2} Q_{\rm total}^T \,\,,
\end{equation}
where
\begin{equation}
Q_{\rm total}^T \equiv -{\mathcal E}_{\rm SB}^{(1)} = \delta_0 \int_{r=0}^{r_B} r^5 dr \epsilon_5 \rho_{T}^{(1)} \,\,.
\end{equation}
$Q_{\rm total}^T $ represents the total vacuum energy (per unit volume of the four-dimensional spacetime) of the brane region which originated from the excitations with components along the transverse directions to the D3-branes. Now in this case the point of the cosmological constant problem can be summarized as whether we can find a nonzero function $\rho_{T}^{(1)}$ satisfying $Q_{\rm total}^T =0$. The existence of such functions implies nonsupersymmetric configurations with vanishing $\lambda$, and important examples of such functions have been found in \cite{1} (see Sec. VII).

\vskip 0.3cm
\hspace{-0.6cm}{(2) With nonvanishing fluxes}
\vskip 0.15cm

Let us now turn on the three-form fluxes $G_{(3)}$ to obtain a full description of the supersymmetry breaking of our AdS vacuum scenario. In the ISD compactifications ($\Phi_{-} = G^{+}_{mnp}=0$), the dilaton $\phi$ still satisfies (8.8) and (8.11) even in the presence of nonzero $G_{(3)}$. However, these ISD compactifications are not appropriate to the general cases of our AdS vacuum scenario because in the AdS minimum, the unbroken supersymmetry $DW=0$ requires that $G_{(3)}$ must also contain IASD $(1,2)$ and $(3,0)$ in addition to the ISD $(2,1)$ and $(0,3)$ (see \cite{9}). These IASD components of the AdS background are entirely due to the nonperturbative corrections of the superpotential and they have nothing to do with the perturbative corrections which also give rise to the IASD components of $G_{(3)}$ and $\Phi_{-}$. In any case, the IASD terms with $G^{+}_{mnp} {\bar G}^{+mnp}$ or $\Phi_{-}$ acquire nonzero values from both perturbative and nonperturbative corrections, and they are now involved $-$ together with those terms caused by $\delta_G {\hat I}_{\rm brane}^{(R)} (=-{\mathcal E}_{\rm SB})$ $-$ in the supersymmetry transformations in some complicated manner. But still, if we ignore the perturbative corrections and supersymmetry breaking generated by $\rho_T^{(1)}$ for a moment, then we expect the solution to the equations of motion becomes a supersymmetric solution satisfying $\delta \chi_{\phi}=\delta \psi_m =0$.\footnote{Supersymmetric solutions of type II theories have been discussed, for instance,  in \cite{23}. (Also see the last paper in \cite{12}.)} Namely the supersymmetry of the AdS background is simply given by $\delta \chi_{\phi}=\delta \psi_m =0$ at the tree level.

Now we finally turn to the situation where the perturbative corrections and supersymmetry breaking generated by ${\mathcal E}_{\rm SB}$ are both taken into account. In this case $\delta \chi_{\phi}$ and $\delta \psi_{m}$ fail to vanish since they now acquire the terms coming from the perturbations and supersymmetry breaking, and consequently the supersymmetries of the brane and bulk regions are both broken. In the case of ${\mathcal V}_{\rm scalar}$, however, the situation is a little different. In the AdS minima of KKLT the scalar potential ${\mathcal V}_{\rm scalar}$ receives contributions both from perturbative and nonperturbative corrections. Hence in this case, ${\mathcal V}_{\rm scalar}$ already takes nonzero values even when perturbations and supersymmetry breakings are not taken into account in the theory yet. However, in our self-tuning mechanism any nonzero contributions to ${\mathcal V}_{\rm scalar}$ coming from perturbative and nonperturbative corrections, and also the contributions coming from the IASD fluxes described above are all gauged away by ${\mathcal E}_{\rm SB}$ in (1.1), and $\lambda =0$ is always preserved even when supersymmetry of the system is broken by the perturbations and supersymmetry breaking term ${\mathcal E}_{\rm SB}$.

In (1.2), we decompose $\rho_T^{(1)}$ into ${\tilde \rho}_T^{(1)} + \delta \rho_T^{(1)}$ to get ${\mathcal E}_{\rm SB} \rightarrow {\tilde {\mathcal E}_{\rm SB}} +\delta {\mathcal E}_{\rm SB}$, where ${\tilde {\mathcal E}_{\rm SB}}$ and $\delta {\mathcal E}_{\rm SB}$ are
\begin{equation}
{\tilde {\mathcal E}_{\rm SB}} = - \delta_0 \int r^5 dr \epsilon_5 {\tilde \rho}_T^{(1)} \,\,, ~~~~~\delta {\mathcal E}_{\rm SB} = -\delta_0 \int r^5 dr \epsilon_5 \delta \rho_T^{(1)} \,\,.
\end{equation}
${\tilde \rho}_T^{(1)}$ and $\delta \rho_T^{(1)}$ in ${\mathcal E}_{\rm SB}$ are arbitrary because they contain six arbitrary gauge parameters $f_m^{(0)}(y)$. Hence if we adjust $\delta \rho_T^{(1)}$ such that $\delta {\mathcal E}_{\rm SB}$ cancels nonzero deviations of ${\mathcal V}_{\rm scalar}$ plus $\delta_Q {\hat I}_{\rm brane}^{(NS)} + \delta_Q {\hat I}_{\rm brane}^{(R)}$ on the brane, then (1.1) reduces to
\begin{equation}
\lambda = \frac{\kappa^2}{2} {\tilde Q}_{\rm total}^T \,\,,
\end{equation}
where ${\tilde Q}_{\rm total}^T$ is the generalized version of (8.15),
\begin{equation}
{\tilde Q}_{\rm total}^T \equiv \delta_0 \int_{r=0}^{r_B} r^5 dr \epsilon_5 {\tilde \rho}_T^{(1)} \,\,.
\end{equation}
The adjustment of $\delta \rho_T^{(1)}$, or the cancelation between $\delta {\mathcal E}_{\rm SB}$ and ${\mathcal V}_{\rm scalar} + \delta_Q {\hat I}_{\rm brane}^{(NS)} + \delta_Q {\hat I}_{\rm brane}^{(R)}$ in (1.1) is automatic by the self-tuning $\lambda =0$ as required by (3.41), and ${\tilde Q}_{\rm total}^T$ now plays the role of $Q_{\rm total}^T$ as one can see from (8.14) and (8.17). So if we want a nonsupersymmetric theory with $\lambda=0$, we may need to find a nonzero function ${\tilde \rho}_T^{(1)}$ satisfying ${\tilde Q}_{\rm total}^T =0$ as in the case of $G_{(3)} =0$. But still, it may also be possible to take simply
\begin{equation}
{\mathcal E}_{\rm SB} = {\mathcal V}_{\rm scalar} + \delta_Q {\hat I}_{\rm brane}^{(NS)} + \delta_Q {\hat I}_{\rm brane}^{(R)} \,\,,
\end{equation}
because this ${\mathcal E}_{\rm SB}$ would be large enough to break the supersymmetry of the system sufficiently.

\vskip 1cm
\hspace{-0.6cm}{\bf \Large Concluding remarks}
\vskip 0.15cm

So far we have considered a new type of self-tuning mechanism to address the cosmological constant problem, especially aiming at explaining the fine-tuning $\lambda =0$ of our present universe. But more precisely, $\lambda$ of our present universe is known to take a positive value though it is very small. So the next step of the project would be this issue of identifying small positive $\lambda$ of our present universe.

In this paper we have considered a theory based on the type IIB supergravity, and from this supergravity action we obtained a result that $\lambda$ must vanish precisely if the density of ${\mathcal V}_{\rm scalar}$ satisfies $V \in V_n$ with $n=1,3$ as in the AdS vacua of our self-tuning mechanism. We have also shown that the result $\lambda=0$ of our AdS vacua is stable against $g_s$-perturbations. But full string theory requires the action to admit $\alpha^{\prime}$-corrections which are usually higher order in derivatives, and due to these corrections the self-tuning equation (3.41) may be modified into the corrected form.

Besides this, in the case of the type I or the heterotic theory the $\alpha^{\prime}$-corrections contain extra terms which do not satisfy $V \in V_n$ with $n=1$ or 3, and these terms also require that $\lambda$ must take nonzero values. Namely if we take the stringy (or any other) effects which have not been considered in this paper into account, we may expect a result with nonvanishing $\lambda$. But still, once $\lambda$ is determined by (the modified) (3.41), these nonzero values of $\lambda$ will be stable against quantum corrections as in the case $\lambda=0$ of this paper because (3.41) is based on the self-tuning mechanism where the perturbative corrections of ${\mathcal V}_{\rm scalar}$ and quantum fluctuations on the branes are always gauged away by ${\mathcal E}_{\rm SB}$ in (1.1). So the result obtained from (3.41) needs to be distinguished from the result of nonvanishing $\lambda$ due to $\alpha^{\prime}$-corrections in the literature \cite{25} in this sense.

In any case, if some convincing values of $\lambda$ is obtained from (3.41) modified by $\alpha^{\prime}$-corrections, then we may say that the nonzeroness of $\lambda$ of our present universe is essentially due to the stringy effect of the string theory, because $\lambda$ vanishes in the absence of $\alpha^{\prime}$-corrections and this result was not affected by the $g_s$-perturbations in our self-tuning mechanism of this paper. But any nonzero values of $\lambda$ suggested by (3.41) will be highly suppressed again by the factor $\chi^{1/2}$ as stated in the last paragraph of Sec. 7.4, and hence $\lambda$ obtained from (3.41) would be very small anyway.

\vskip 1cm
\begin{center}
{\large \bf Acknowledgement}
\end{center}

This work was supported by the National Research Foundation of Korea (NRF), under Grant No. 353-2009-2-C00045, funded by the Korean Government.

\end{document}